\documentclass[acmsmall, screen,nonacm]{acmart}

\usepackage[USenglish]{babel}

\usepackage{stmaryrd}

\usepackage{graphicx}
\usepackage{paralist}
\usepackage{caption}
\usepackage{subcaption}

\usepackage{tikz}
\usetikzlibrary{arrows, arrows.meta, automata, positioning, calc}

\usepackage{bussproofs}
\usepackage{algorithm2e}
\usepackage{wrapfig}
\usepackage{listings}
\usepackage[most]{tcolorbox}

\newcommand{\rayna}[1]{\textcolor{magenta}{}}

\newcommand{\best}[1]{\textbf{#1}}
\newcommand{\vbest}[1]{\underline{\textbf{#1}}}

\newcommand{\Nat}{\ensuremath{\mathbb{N}}}
\newcommand{\Bool}{\ensuremath{\mathbb{B}}}
\newcommand{\Int}{\ensuremath{\mathbb{Z}}}

\newcommand{\Real}{\ensuremath{\mathbb{R}}}

\newcommand{\power}[1]{\ensuremath{2^{#1}}}
\newcommand{\sema}[1]{\llbracket #1 \rrbracket}

\newcommand{\values}{\ensuremath{\mathcal{V}}}
\newcommand{\functions}{\ensuremath{\mathcal{F}}}
\newcommand{\predicates}{\ensuremath{\mathcal{P}}}

\newcommand{\vars}{\ensuremath{\mathit{Vars}}}

\newcommand{\funcSymbols}{\ensuremath{\Sigma_F}}
\newcommand{\predSymbols}{\ensuremath{\Sigma_P}}
\newcommand{\funcTerms}{\ensuremath{\mathcal{T}_F}}
\newcommand{\predTerms}{\ensuremath{\mathcal{T}_P}}

\newcommand{\assignment}{\ensuremath{\nu}}
\newcommand{\assignments}[1]{\ensuremath{\mathit{Assignments}}(#1)}
\newcommand{\interpretation}{\ensuremath{\mathcal I}}
\newcommand{\interpretations}[1]{\ensuremath{\mathit{Interpretations}}(#1)}

\newcommand{\eval}[1]{\ensuremath{\chi_{#1}}}

\newcommand{\assmt}[1]{\mathbf{#1}}

\newcommand{\FOL}[1]{\ensuremath{\mathit{FOL}}(#1)}
\newcommand{\FOLX}{\ensuremath{\mathit{FOL}}}
\newcommand{\QF}[1]{\ensuremath{\mathit{QF}}(#1)}
\newcommand{\QFX}{\ensuremath{\mathit{QF}}}
\newcommand{\FOLentails}{\ensuremath{\models}}
\newcommand{\FOLentailsT}[1]{\ensuremath{\models_{#1}}}

\newcommand{\valid}[1]{\textsc{Valid}(#1)}

\newcommand{\cells}{\ensuremath{\mathbb{X}}}
\newcommand{\inputs}{\ensuremath{\mathbb{I}}}

\newcommand{\guards}{\ensuremath{\mathit{Guards}}}
\newcommand{\updates}{\ensuremath{\mathit{Updates}}}
\newcommand{\inputa}{\ensuremath{\mathit{Inputs}}}

\newcommand{\states}{\ensuremath{\mathcal{S}}}
\newcommand{\symstates}{\ensuremath{\mathcal{D}}}

\newcommand{\cpre}[2]{\ensuremath{\mathit{CPre}_{#1}(#2)}}

\newcommand{\attr}[2]{\ensuremath{\mathit{Attr}_{#1}(#2)}}

\newcommand{\sys}{\ensuremath{\mathit{Sys}}}
\newcommand{\env}{\ensuremath{\mathit{Env}}}
\newcommand{\s}{\ensuremath{\mathsf{s}}}
\newcommand{\e}{\ensuremath{\mathsf{e}}}

\newcommand{\loc}{\ensuremath{\mathit{loc}}}

\newcommand{\enabled}[1]{\mathit{Act}_{#1}}
\newcommand{\str}[2]{\mathit{Strat}_{#1}(#2)}
\newcommand{\outcome}[3]{\mathit{Outcome}(#1,#2,#3)}
\newcommand{\plays}{\mathit{Plays}}

\newcommand{\reach}{\mathit{Reach}}
\newcommand{\safe}{\mathit{Safety}}
\newcommand{\buchi}{\mathit{Buchi}}
\newcommand{\cobuchi}{\mathit{coBuchi}}

\newcommand{\changed}[1]{\textcolor{blue}{#1}}

\newcommand{\Inv}{\ensuremath{\mathit{Inv}}}
\newcommand{\inv}{\ensuremath{\mathit{inv}}}
\newcommand{\base}{\ensuremath{\mathit{base}}}
\newcommand{\conc}{\ensuremath{\mathit{conc}}}
\newcommand{\step}{\ensuremath{\mathit{step}}}

\newcommand{\loopgame}{\ensuremath{\mathsf{LoopGame}}}
\newcommand{\constraints}{\ensuremath{\mathit{Constraints}}}
\newcommand{\constr}{\ensuremath{\Psi}}
\newcommand{\accreach}{\ensuremath{\mathsf{AccA}}}
\newcommand{\acccobuchi}{\ensuremath{\mathsf{AccB}}}
\newcommand{\lemmasymb}{\ensuremath{\mathsf{LemmaSymb}}}
\newcommand{\usedlemmas}{\ensuremath{\mathsf{UsedLemmaSymbols}}}
\newcommand{\lemmas}{\ensuremath{\mathsf{Lemmas}}}
\newcommand{\itera}{\ensuremath{\mathsf{IterA}}}
\newcommand{\iterb}{\ensuremath{\mathsf{IterB}}}

\newcommand{\precise}{\ensuremath{\mathsf{Precise}}}
\newcommand{\resolve}{\ensuremath{\mathsf{InstantiateLemmas}}}
\newcommand{\cellsnul}{E}

\newcommand{\locsplit}{\ensuremath{l_\mathit{Split}}}
\newcommand{\locend}{\ensuremath{l_\mathit{End}}}
\newcommand{\combine}[2]{\langle #1, #2 \rangle}

\renewcommand{\cite}[1]{\citep{#1}}

\begin{document}

\title{Solving Infinite-State Games via Acceleration (Full Version)}

\author{Philippe Heim}
\email{philippe.heim@cispa.de}
\orcid{0000-0002-5433-8133}
\affiliation{%
  \institution{CISPA Helmholtz Center for Information Security}
  \streetaddress{Stuhlsatzenhaus 5}
  \city{Saarbrücken}
  \country{Germany}
  \postcode{66123}
}

\author{Rayna Dimitrova}
\email{dimitrova@cispa.de}
\affiliation{%
  \institution{CISPA Helmholtz Center for Information Security}
  \streetaddress{Stuhlsatzenhaus 5}
  \city{Saarbrücken}
  \country{Germany}
  \postcode{66123}
}

\ccsdesc[500]{Theory of computation~Logic and verification}
\ccsdesc[500]{Software and its engineering~Automatic programming}

\keywords{infinite-duration games, infinite-state games, reactive synthesis}

\begin{abstract}

Two-player graph games have found numerous applications,  most notably in the synthesis of reactive systems from temporal specifications, but also in verification.
The relevance of infinite-state systems in these areas has lead to significant attention towards developing techniques for solving infinite-state games.

We propose novel  symbolic semi-algorithms for solving infinite-state games with temporal winning conditions.
The novelty of our approach lies in the introduction of an acceleration technique that enhances fixpoint-based game-solving methods and helps to avoid divergence.
Classical fixpoint-based algorithms, when applied to infinite-state games,  are bound to diverge in many cases, since they iteratively compute the set of states from which one player has a winning strategy. 
Our proposed approach can lead to convergence in cases where existing algorithms require an infinite number of iterations.
This is achieved by acceleration: computing an infinite set of states from which a simpler sub-strategy can be iterated an unbounded number of times in order to win the game.
Ours is the  first method for solving infinite-state games to employ acceleration.
Thanks to this,  it is able to outperform state-of-the-art techniques on a range of benchmarks,  as evidenced by our evaluation of a prototype implementation.

\end{abstract}

\maketitle

\section{Introduction}\label{sec:intro}
Reactive synthesis, introduced by Church~\cite{Church62}, has the goal of automatically generating an implementation (for instance, a program or a finite-state controller) from a formal specification that describes the desired behavior of the system.
The system requirements are typically specified using temporal logics, such as Linear Temporal Logic (LTL), which provide expressive high-level specification languages.
Recent advancements~\cite{BloemJPPS12,FinkbeinerS13} have made possible the successful application of synthesis to industrial protocols and robotics.

The problem of synthesizing strategies in two-player games over graphs is tightly connected to reactive synthesis.
Synthesis from temporal logic specifications can be reduced to computing a winning strategy in a game.
Games can also be used to model the interaction between a system and its environment directly.
There is a large body of algorithmic techniques and tools for solving finite-state games.
Many applications of two-player games, however,  naturally require the treatment of infinite-state models, such as software synthesis and repair~\cite{GriesmayerBC06},  controller synthesis in domains like robotics~\cite{Kress-GazitFP09},  and software verification against hyperproperties~\cite{BeutnerF22}.
The problem of solving (i.e., determining the winner) infinite-state games is in general undecidable, and many practical applications lie outside of decidable classes, making incomplete approaches necessary.
As different approaches have different strengths, a number of techniques have been developed over the last years~\cite{BeyeneCPR14, NeiderT16, FarzanK18, SamuelDK21}.
However, the state of the art is still far from the level of the algorithmic approaches and tools available for the finite-state case.

We propose a novel technique for solving infinite-state games that aims to address one of the limitations of existing approaches, namely, that they usually diverge on game-solving tasks that require reasoning about the unbounded iteration of strategic decisions.
We illustrate this challenge with an example.

\begin{figure}[t!]
 \begin{subfigure}[b]{0.56\textwidth}
  \begin{center}
  \begin{tikzpicture}[->,>=stealth',shorten >=1pt,auto,node distance=2.5cm]
  \tikzstyle{every state}=[fill=none,draw=black,text=black,inner sep=1.5pt, minimum size=16pt,thick,scale=0.7]
    \node[state] (l) at (0,0) {$l_0$};
    \node[state] (lG) at (4,0) {$l_G$};

    \node[fill=black,draw=black,minimum size=0.5pt] (e1) at (-2.5,0) {};
    \node[fill=black,draw=black,minimum size=0.5pt] (e2) at (2.5,0) {};

    \draw (l) -- node[above] {\small $x > 42 \textcolor{blue}{\land i \neq 0}$} (e1);
    \draw[rounded corners] 
        (e1) -- ++(0,0.7) -- node[above] {\small $x := x + i$} ++(2.5,0) -- (l);
    \draw[rounded corners] 
        (e1) -- ++(0,-0.7) -- node[above] {\small $x := x - i$} ++(2.5,0) -- (l);
    \draw (l) -- node[above] {\small $x \leq 42 \textcolor{blue}{\lor i = 0}$} (e2);
    \draw (e2) -- node[above] {\small $x := x$} (lG);
    \draw[rounded corners] 
        (lG) -- ++(0,0.7) -- node[above] {\small $\top$} ++ (-1.5,0) -- (e2); 
\end{tikzpicture}
  \end{center}
    \caption{Reactive program game structure with two locations ($l_0$ and $l_G$) and two integer variables: environment-controlled \emph{input variable $i$} and system-controlled \emph{program variable} $x$.  Edges to black squares are labeled with \emph{guards} (formulas over the variables),   and edges originating in black squares are labeled with  \emph{updates} to the program variables.  }\label{fig:ex-motivating}
\end{subfigure}\hfill
 \begin{subfigure}[b]{0.41\textwidth}
  \begin{center}
    \scalebox{0.8}{\input{example-program}}
  \end{center}
    \caption{Reactive program with input variable $i$, program variable $x$ and  auxiliary variable $x_0$.}
    \vspace{-.3cm}
    \label{fig:ex-program}
\end{subfigure}    
\caption{Reactive program game (left) and a corresponding synthesized reactive program (right).}
\vspace{-5mm}
\end{figure}

\begin{example}\label{ex:intro}
Consider the simple game shown in Figure~\ref{fig:ex-motivating}.
It models the interaction between a reactive program (the system) and its environment.
The game has two locations, $l_0$ and $l_G$, an integer input variable $i$, and an integer program variable $x$. 
The edges depict transitions and are labeled with guard conditions over the input and output variables,  and updates which are  assignments to the program variable.
When we depict transitions we separate guards and updates.  
Edges originating in locations are labeled with guards and end in a black square.
Edges originating in the black squares are labeled with possible updates for the system to choose from.
Note that the black squares are used simply for visualization. 
When playing the game, in each step the environment chooses a value for $i$.
Then, in $l_0$, if $x$ is smaller than 42 or $i$ is zero, the system must leave $x$ unchanged and the game moves to $l_G$.  
Otherwise, the system has to decide whether to add or subtract $i$ from $x$ and the game stays in $l_0$.
Once in $l_G$, the game stays there forever and $x$ is not changed.

Let us consider the \emph{temporal specification} where the system is required to \emph{eventually reach} location $l_G$ from $l_0$ for any initial value in $x$.
The program can indeed enforce this property by \emph{choosing the appropriate update at every step}: adding $i$ to $x$ if $i < 0$ and subtracting $i$ from $x$ if $i > 0$.
This ensures that $x$ is decremented in every step (unless the environment sets $i$ to $0$, in which case the game moves to $l_G$), and hence $ i = 0 \vee x \leq 42$ will eventually be reached.

The number of times $x$ has to be decremented depends on its initial value, and therefore \emph{the number of iterations through $l_0$ is a priori unbounded}.
However, not only the number of iterations in the system execution is unbounded, which would correspond to an unbounded loop, but in every iteration, the system has to perform \emph{strategic decisions} which depend on the input from the environment and impact the execution.
We call these constructs \emph{unbounded strategy loops}.

While necessary for many practical applications, handling unbounded strategy loops is out of the scope of existing methods for reactive synthesis.
However,  in our example, the necessary argument to establish that the system can satisfy the requirement of the game can be stated as the following simple property:
{\bf Whenever $x > 42$, if the system can ensure that it can decrement $x$, then eventually $x \leq 42$ will be reached}.
To utilize this reasoning, the synthesis process has to establish that the task postulated in the sub-property, i.e.\ ``$x$ is decremented'', can be realized in certain situations, i.e.\ $i \neq 0$.
This itself is a reactive synthesis task.
In this paper we propose a symbolic method for solving infinite-state games that utilizes such  reasoning in order to handle unbounded strategy loops as the one in Figure~\ref{fig:ex-motivating}.
\qed
\end{example}

\paragraph{Contributions}
We present a symbolic approach for solving infinite-state games with temporal objectives, such as reachability,  safety,  Büchi, and co-Büchi,  lifting the respective finite-state algorithms.
The key novelty of our technique is an \emph{acceleration method} which, in contrast to existing approaches,  accelerates unbounded strategy loops.
It uses the newly introduced notion of \emph{acceleration lemmas},  which are simple inductive statements that the symbolic algorithm lifts to accelerate the termination of the game-solving process.
To the best of our knowledge,  this is the first acceleration-based method for game solving.
We implemented our method in a prototype and demonstrate its feasibility by its successful application to standard benchmarks from the literature, outperforming existing tools, and several new ones that are out of the scope of these tools.

\paragraph{Paper Outline}

After discussing related work in Section~\ref{sec:related}, we present in Section~\ref{sec:examples} several examples illustrating different challenges, how our technique approaches them, and where existing methods fall short. 
After the technical preliminaries in Section~\ref{sec:prelim}, we introduce our game model in Section~\ref{sec:games}. 
We then describe the basis of our solving techniques in Section~\ref{sec:solving}. 
In Section~\ref{sec:acceleration}, we present formally our acceleration method.
We continue by discussing the synthesis of acceleration lemmas in Section~\ref{sec:resolving}.
Lastly,  Section~\ref{sec:eval} evaluates our method and discusses our empirical observations.
\looseness=-1

\section{Related Work}\label{sec:related}
We now overview the landscape of existing approaches for solving infinite-state games and explain how our work addresses some of their limitations. 
We also discuss relevant techniques for verification of infinite-state systems and why they do not directly apply in the context of synthesis.

\paragraph{Infinite-State Game Solving.}

In general, games over infinite graphs are undecidable.
Decidable classes, such as pushdown games~\cite{Walukiewicz01} and downward-closed safety games~\cite{AbdullaBd08}, as well as termination criteria for symbolic procedures~\cite{AlfaroHM01} have been found.
As many practical applications lie outside of these classes, incomplete methods are needed.

Different approaches have been proposed for safety games.
\cite{NeiderT16} presents an automata-learning method for safety games over infinite graphs, 
and~\cite{MarkgrafHLNN20} develops a learning-based technique for parameterized systems with safety specifications.  
In~\cite{FarzanK18}, the authors study safety and reachability games defined within the theory of linear rational arithmetic.
\cite{KatisFGGBGW18} presents a method for synthesis from Assume-Guarantee contracts describing safety properties.
The tool GenSys~\cite{SamuelDK21} implements a fixpoint-based solver for infinite-state safety games using an SMT solver.
\cite{FaellaP23} presents a method for solving infinite-state reachability games by reducing the problem to checking the satisfiability of a system of constrained Horn clauses.  
It is restricted to games where the reachability player has finitely many actions to choose from.
More expressive winning conditions,  such as the Büchi condition which requires repeated visits to some states,  are out of the scope of these methods.

The constraint-based approach in~\cite{BeyeneCPR14} handles infinite-state games with winning conditions given by LTL specifications.  
It can solve problems with unbounded strategy loops.
However, 
it requires the user to provide  
templates that structure how the final system works, including handling the unbounded strategy loops. 
As can be seen in the examples in~\cite{BeyeneCPR14} such templates can be quite  complex even for small games.
In contrast, our approach uses inductive statements that are automatically generated from generalizable templates.

Solvers for first-order fixpoint logics~\cite{UnnoTGK23,UnnoSTK20} can be used, as described in~\cite{UnnoSTK20},  to solve games with omega-regular winning conditions. 
This approach uses the fixpoint encoding of the winning sets of states for a player in the game.  
Its ability to find a solution depends on the constraint-solving engine.  
In Section~\ref{sec:eval} we report on experimental comparison with the technique described in~\cite{UnnoSTK20}, demonstrating the strengths of our approach.

Abstraction-based techniques have been extended to games~\cite{HenzingerJM03,GrumbergLLS05,BallK06,GrumbergLLS07,WalkerR14}.  
Abstraction-based controller synthesis for dynamical systems~\cite{Tabuada09} makes use of discretization.
As they reduce the solving of an infinite-state game to the finite-state case, they can use the techniques and tools available for finite-state games. 
The effectiveness of these approaches depends on the abstractions,  whose iterative refinement might diverge when unbounded strategy loops are needed.

\paragraph{Infinite-State Synthesis from Temporal Logics.}
The recent works~\cite{ChoiFPS22,MaderbacherB21} both study the problem of synthesizing reactive systems from Temporal Stream Logic modulo theories (TSL-MT~\cite{FinkbeinerHP22}) specifications. 
This temporal logic can express conditions on unbounded input data and program variables.
\cite{ChoiFPS22} integrates a procedure for synthesis from TSL specifications~\cite{Finkbeiner0PS19} and Syntax-Guided synthesis.
The role of SyGuS is to generate assumptions that are added to a refined TSL formula.
The procedure proposed in~\cite{MaderbacherB21} is based on abstraction refinement and performs a counterexample-guided synthesis loop that invokes LTL synthesis.  
The refinement uses an SMT solver to analyze counterstrategies for inconsistency with the theory.
While \cite{ChoiFPS22} can handle some unbounded looping behavior by using recursive functions, both~\cite{MaderbacherB21, ChoiFPS22} cannot handle unbounded strategy loops where at each iteration, the environment provides new input values.
\cite{SamuelDK23} proposes a symbolic fixpoint computation for infinite-state games with LTL winning conditions.  As the authors remark,  their method has the same non-termination issues as~\cite{MaderbacherB21}, which implies that it will also diverge in the presence of unbounded strategy loops.

\paragraph{Infinite-State Verification.} 
There is a vast variety of approaches for verification of infinite-state systems. 
Above we discussed extensions of abstraction-based techniques and deductive verification techniques to the setting of games and synthesis of reactive systems.  
In contrast, other prominent approaches used in the verification of infinite-state systems, such as acceleration~\cite{FinkelL02,BardinFLP03,  BardinFLS05, LerouxS06} or loop summarization~\cite{KroeningSTTW13}, do not directly extend to the setting of two-player games, and have thus far not been explored. 
The key difficulty is caused by the alternation of environment inputs and decisions by the system player,  meaning that a loop in the game structure might not be under the full control of the system.  Thus, the existence of a loop whose transitions can be composed, does not entail that it is enforceable by the system.  
We illustrate this in the next section, and discuss this challenge further in this paper.

\section{Overview and Motivating Examples}\label{sec:examples}
In this section, we give a high-level overview of our approach, highlighting some of its strengths and distinguishing features on several simple but challenging examples.
First,  we show how our acceleration method can be applied to enable termination in certain cases, specifically in the presence of unbounded strategy loops ( Example~\ref{ex:intro}), where existing techniques typically diverge.  
We explain, on the high level,  how acceleration works,  and also discuss what are the challenges that our technique addresses. 
Second,  we demonstrate the advantage of our acceleration-based game solving procedure over purely constraint-based approaches for solving fixpoint equations.  
Finally,  we illustrate the applicability of our acceleration method to expressive classes of specifications encoded as B\"uchi objectives in the synthesis game.

We begin with a brief,  high-level,  description of our game model and problem formulation. 
The underlying  formal definitions  are presented in the later Section~\ref{sec:prelim} and  Section~\ref{sec:games}.

\subsection{Symbolic Model and Problem Formulation}

We specify synthesis tasks as \emph{reactive program game structures}, like the one depicted in Figure~\ref{fig:ex-motivating}.  
Such a game structure consists of a finite set $L$ of control locations, a finite set $\inputs$ of input variables, and a finite set $\cells$ of output variables.  While the sets $\inputs$ and $\cells$ are finite,  the domains of the variables can be infinite. Thus, reactive program game structures can represent infinite-state games.  A reactive program game contains transitions between the locations, labeled with guards and updates.   A transition has the form $(l,g,u,l')$, where $l$ and $l'$ are the source and target locations respectively,   $g$ is the guard,  and $u$ is the update.  As in Figure~\ref{fig:ex-motivating}, we depict such a transition using an edge from $l$ to a black square labeled with $g$, and an edge from a black square to $l'$ labelled with $u$.  The intermediate black squares are for visualization only, and not part of the formal definition.
A guard $g$ has to hold for the transition to be possible, and an update $u$ assigns to each program variable a term with whose value it should be updated. 
A \emph{reactive program game} is pair of a reactive program game structure and a \emph{winning condition} expressed over the locations $L$.  For example,  in Figure~\ref{fig:ex-motivating},  we  consider the reachability condition that requires that location $l_G$ is eventually visited. 
In Example~\ref{ex:buchi-motivating} we show a reactive program game with a Büchi winning condition that states that the system has to enforce visiting a set of accepting locations $B \subseteq L$ infinitely often.

The reactive program game is played over the possibly infinite set of states,  where each state $(l,\assignment)$ consists of the current location $l$ and an assignment $\assignment$ of values for all the program variables $\cells$. 
A step in the game is played as follows.
In the current state $(l,\assignment)$,  the environment chooses some values $\assmt{i}$ for the inputs $\inputs$. 
Then, the system chooses a transition $(l,g,u,l')$ whose guard $g$ is satisfied by the current values $\assmt{i}$ and $\assignment$, of the inputs and the program variables respectively. 
The next location is $l'$,  and the next values of the program variables are determined according to the  update $u$.  
Starting in some state $(l,\assignment)$,  this repeated interaction produces a play,  an infinite sequence of states,  which is winning for the system player if it satisfies the given winning condition.  

The \emph{realizability problem} is to determine whether there exists a strategy for the system to resolve the transition (i.e., update) choices that guarantees that any play starting in any possible state $(l_{\mathit{init}},\assignment)$ for a given initial location $l_{\mathit{init}}$,  is winning for the system, regardless of the  input values chosen by the environment.  
The \emph{synthesis problem} asks to compute such a strategy if one exists.  One such a strategy for Example~\ref{ex:intro} with initial location $l_0$ is depicted in Figure~\ref{fig:ex-program}, in the form of a reactive program. The program contains an auxiliary variable $x_0$, whose role will be explained later.

\subsection{Acceleration for Unbounded Strategy Loops}

Finite-state games can be solved by fixpoint algorithms that compute sets of states from which a given  player can enforce winning the game.
To solve reactive program games, we lift those algorithms to compute symbolically infinite sets of states. 
We represent them as elements of a symbolic domain $\symstates := L \to \FOL{\cells}$, 
where $\FOL{\cells}$ is the set of first-order logic formulas with free variables among the program variables $\cells$.
A symbolic set $d \in \symstates$ represents the set of those $(l,\assignment)$ where the assignment $\assignment$ to $\cells$ satisfies $d(l)$.
A basic notion in the fixpoint-based game-solving algorithms is the \emph{attractor}. 
For a given player and a given set of goal states,  the attractor consists of the states from which the player can enforce reaching a goal state in some number of steps.  Attractor sets can be computed via iterative fixpoint computation, which is not guaranteed to terminate in the case of an infinite state space.  We illustrate this on Example~\ref{ex:intro}.

\textit{Example 1.1 (Continued).}
Suppose we want to compute the attractor for the system player for the symbolically represented set of states $d := \{ l_0 \mapsto \bot, l_G \mapsto \top \}$ (describing the set of states whose location is $l_G$).
After the first iteration, we get $\{ l_0 \mapsto (x \leq 42), l_G \mapsto \top \}$. 
In the next one, we get $\{ l_0 \mapsto (\forall i. x \leq 42 \lor i = 0 \lor x + i \leq 42 \lor x - i \leq 42),  l_G \mapsto \top \}$  which simplifies to $\{ l_0 \mapsto (x \leq 43),  l_G \mapsto \top \}$.
Then $\{l_0 \mapsto (x \leq 44),  l_G \mapsto \top\}$, $\{l_0 \mapsto (x \leq 45),  l_G \mapsto \top\}$,  \dots{}.  A fixpoint is never reached, since at every step of the computation we add one state from the infinite attractor set. \looseness=-1
\qed

In Section~\ref{sec:intro} we argued intuitively why in Example~\ref{ex:intro} every state $(l_0, \nu)$ belongs to the attractor for the system player.  The method we propose in this paper is able to automatically establish this and compute the attractor.
More concretely,  this is achieved as follows.
\begin{itemize}
\item We introduce the notion of \emph{acceleration lemmas}, which allow us to express in a formal way arguments like the one outlined in Section~\ref{sec:intro}.   
An acceleration lemma, precisely defined in Section~\ref{sec:acceleration-attractor}, is a triple $(\base, \step, \conc)$ of \FOLX\ formulas.  The \emph{conclusion} $\conc$ characterizes a set of states with the property  that from each of them, by iterating  the \emph{step relation} described by $\step$, a state in the set characterized by the base condition $\base$ must be reached.  Intuitively, $\conc$ characterizes the states added to the attractor by applying this acceleration lemma.
For our Example~\ref{ex:intro},  one possible acceleration lemma is $(\base,\step,\conc)$ with $\base := (x \leq 42)$,  $\step := (x' < x)$ and $\conc := \top$.
In the formula $\step := (x' < x)$,  the variable $x$ refers to the value of the respective program variable at one visit to location $l_0$ and $x'$ refers to its value the next time $l_0$ is visited.  Clearly,  starting in any state $(l_0,\nu_{0})$,  any sequence of states $(l_0,\nu_{0}),(l_0,\nu_{1}),\ldots$ that satisfies $\step$ eventually reaches a state that satisfies $x \leq 42$.

\item The purpose of an acceleration lemma $(\base,\step,\conc)$ is to accelerate the computation of an attractor set for a given player $p$ in a reactive program game structure $\mathcal G$, 
by adding all the states  $(l, \assignment)$ that satisfy $\conc$ to the attractor at once. 
In order to apply such a lemma,  the attractor computation procedure must first ensure that the set of states  described by  $\base$ is included in the subset of the attractor computed thus far.  
Second,  it is necessary to establish that player $p$ can \emph{enforce the repeated iteration of the step relation} in the game $\mathcal G$ \emph{against all possible behaviors of the opponent player}.
Intuitively,  this can be guaranteed by showing that player $p$ has a strategy to enforce looping in location $l$ such that each iteration satisfies $\step$.
In Example~\ref{ex:intro} this is indeed the case for location $l_0$ and the acceleration lemma given above. 
Thus, our method is able to add all states $(l_0,\nu)$ to the attractor for the system player. 
\end{itemize}
 
In Section~\ref{sec:acceleration} we describe in detail how our method generates acceleration lemmas while at the same time ensuring that their step relations can be enforced by the respective player in the underlying reactive program game structure.  Next,  we give an example that illustrates why the latter condition is crucial for the correct application of an acceleration lemma.  In fact, this is one of the major challenges for acceleration in the context of synthesis, and a key difference to the application of acceleration techniques in verification.

\begin{example}\label{ex:unrealizable}
Consider a modification of the reactive program game depicted in Figure~\ref{fig:ex-motivating} in which we have omitted the parts of the guards referring to the input variable $i$ and colored in blue.  
Clearly,  now the only states $(l_0,\nu)$ for which the system can enforce reaching location $l_G$ are those where $x \leq 42$,  since otherwise the environment can prevent the system from decreasing  $x$ by always setting $i$ to $0$. 
Thus, although the reactive program game contains loops that decrement $x$,   \emph{these loops cannot be enforced by the system}.  This is a major difference to the acceleration techniques employed in verification,  as the existence of a loop with certain properties is a realizability problem in itself.  In Section~\ref{sec:acceleration-attractor} we show how we address this problem by the utilization of a so-called \emph{loop game}.\looseness=-1
\qed
\end{example}

\subsection{Embedding Acceleration in Iterative Symbolic Fixpoint Computation}
Our acceleration technique is embedded in symbolic game-solving algorithms based on iterative fixpoint computation.  This enables our method to combine the strengths of both techniques by utilizing acceleration for unbounded strategy loops in concert with performing a finite  number of concrete iterations of the fixpoint computation. 
The latter can be helpful when the set of winning states that the game solving procedure has to compute  does not have a  ``simple'' and ``easy to discover'' symbolic characterization, as the next example illustrates.

\begin{figure}[t!]
  \begin{center}
  \scalebox{0.8}{\begin{tikzpicture}[->,>=stealth',shorten >=1pt,auto,node distance=2.5cm]
  \tikzstyle{every state}=[fill=none,draw=black,text=black,inner sep=1.5pt, minimum size=16pt,thick,scale=0.8]
    \node[state] (l0) at (0,0) {$l_0$};
    \node[state] (l1) at (5,0) {$l_1$};    
    \node[state] (lg) at (10,0) {$l_G$};
    \node[state] (lb) at (10,-1.3) {$l_B$};

    \node[fill=black,draw=black,minimum size=0.5pt] (e01) at (2.5,0) {};
    \node[fill=black,draw=black,minimum size=0.5pt] (e1g) at (7.5,0) {};
    \node[fill=black,draw=black,minimum size=0.5pt] (e1b) at (7.5,-1.3) {};
    \node[fill=black,draw=black,minimum size=0.5pt] (e11) at (5,1.5) {};

    \draw (l0) -- node[below,align=left] {\small $\top$} (e01); 
    \draw (e01) -- node[below,align=left] {\small $x:=1$} (l1); 
    \draw (l1) -- node[above, near end] {\small $x = 64$} (e1g);
    \draw[rounded corners] 
        (l1) -- ++ (0,-1.3) -- node[above, align=left] {\small $x \neq 64 \wedge $\\$ y \leq 0$} (e1b);
    \draw (e1g) -- node[below,align=left] {\small $\mathsf{skip}$} (lg);
    \draw (e1b) -- node[below,align=left] {\small $\mathsf{skip}$} (lb);
	\draw[rounded corners] 
        (e01) -- ++(0,0.5) -- node[above,align=left] {\small $y:=y+1$} ++(-2.5,0) -- (l0);
	\draw[rounded corners] 
        (lg) -- ++(0,0.4) -- node[above,align=left] {\small $\top$} ++(-2.5,0) -- (e1g);
	\draw[rounded corners] 
        (lb) -- ++(0,0.4) -- node[above,align=left] {\small $\top$} ++(-2.5,0) -- (e1b);
        
   \path
 	(l1) edge[bend left=25] node[left, align=left] {\small $x \neq 64 \wedge $\\$ y >0$} (e11)
    (e11) edge[bend left=25] node[right,align=left] {\small $x:=2 \cdot x;$\\$y:=y-1$} (l1)
    ;

\end{tikzpicture}}
  \end{center}
  \vspace{-3mm}
  \caption{Reactive program game structure with two system-controlled variables $x$ and $y$ and no input variables. We use  $\mathsf{skip}$ to denote the update $x := x; y := y$.  If the update to some program variable is missing from the label of an update edge,  this means that the value of this variable remains unchanged.  }
\label{fig:ex-combined}
  \vspace{-3mm}
\end{figure}

\begin{example}\label{ex:combined}
Consider the reactive program game structure depicted in Figure~\ref{fig:ex-combined}. 
The specification again requires the system to reach location $l_G$ from $l_0$ for any initial values of the program variables $x$ and $y$.
Note that if the value of $y$ is at least $6$ when transitioning from location $l_0$ to $l_1$,  the system can reach $l_G$ after doubling $6$ times the value of $x$. Regardless of the initial values of the variables in $l_0$, it is possible to reach $l_1$ with $y$ having value $6$ or larger. 
Thus, the specification is realizable.

If linear arithmetic formulas are used to represent sets of states,  then the set of values at location $l_1$ from which $l_G$ can be reached is characterized by the formula $(x=64) \vee (x = 32 \land y \geq 1) \vee (x = 16 \land y \geq 2)\vee (x = 8 \land y \geq 3)\vee (x = 4 \land y \geq 4)\vee (x = 2 \land y \geq 5)\vee (x = 1 \land y \geq 6)$.  Constraint-solving techniques for fixpoint equation systems can have difficulty generating such formulas, as we will see in our experimental evaluation in Section~\ref{sec:eval}. 
On the other hand,  methods for iterative symbolic attractor computation that do not apply acceleration diverge due to the presence of the loop in $l_0$. 

Our acceleration-based procedure successfully determines the realizability  in this case, as it  integrates acceleration in the iterative fixpoint computation.
Note that, for simplicity, this example does not contain environment inputs, but the same challenges are present even if it does. 
\qed
\end{example}

\subsection{Acceleration Beyond Reachability and Safety Games}

The synthesis approach that we propose is applicable beyond reachability specifications.  For instance,  we consider Büchi specifications that require that some set of locations is visited infinitely often,  or co-Büchi specifications that require that some set of locations is visited only finitely many times.  
Accelerated attractor computation can be readily integrated in symbolic fixpoint-based procedures for infinte-state games with these types of objectives.
The next example shows a reactive program game with a Büchi specification,  for which our attractor acceleration method allows us to establish realizability.

\begin{figure}[b!]
  \vspace{-3mm}
  \begin{center}
  \scalebox{0.8}{\begin{tikzpicture}[->,>=stealth',shorten >=1pt,auto,node distance=2.5cm]
  \tikzstyle{every state}=[fill=none,draw=black,text=black,inner sep=1.5pt, minimum size=16pt,thick,scale=0.8]
	\node[state] (l0) at (3,0) {$l_0$};    
    \node[state] (l1) at (8,0) {$l_1$};
    \node[state,accepting] (l2) at (4.5,1) {$l_2$};
	\node[state,accepting] (l3) at (1,-1) {$l_3$};
    \node[state] (l4) at (-1,0) {$l_4$}; 

    \node[fill=black,draw=black,minimum size=0.5pt] (e01) at (6,0) {};
    \node[fill=black,draw=black,minimum size=0.5pt] (e11a) at (13,1) {};
    \node[fill=black,draw=black,minimum size=0.5pt] (e11b) at (13,-1) {};
    \node[fill=black,draw=black,minimum size=0.5pt] (e20) at (3,1) {};
    \node[fill=black,draw=black,minimum size=0.5pt] (e30) at (1,0) {};
	\node[fill=black,draw=black,minimum size=0.5pt] (e12) at (5.7,1) {};
    \node[fill=black,draw=black,minimum size=0.5pt] (e34) at (-1,-1) {};

   \draw (l0) -- node[below] {\small $\top$} (e01); 
   \draw (e01) -- node[below] {\small $x:= i$} (l1);  
   \draw[rounded corners] 
        (e01) -- ++(0,-1) -- node[above] {\small $y:=y-1$} (l3);    
   \draw[rounded corners] 
        (l1) -- ++ (0,1) -- node[above, align=right] {\small $x \leq 42 \lor i = 0$} (e12);
   \draw (e12) -- node[above] {\small $\mathsf{skip}$} (l2);
   \draw (l2) -- node[above] {\small $\top$} (e20);
   \draw (e20) -- node[right] {$y:=64$} (l0);
   \draw (l3) -- node[right] {\small $y \geq 16$} (e30);
   \draw (e30) -- node[above] {\small $\mathsf{skip}$} (l0);  
   \draw (l3) -- node[above] {\small $y < 16$} (e34);
   \draw (e34) -- node[left] {\small $\mathsf{skip}$} (l4);  
   \draw[rounded corners] 
        (l4) -- ++(-1,0) -- node[left] {\small $\top$} ++(0,-1) -- (e34);

   \draw[rounded corners]
        (l1) -- ++ (1,1) -- node[above] {\small $x > 42 \land i \neq 0 \land y \leq 32$} (e11a);
   \draw[rounded corners] 
        (e11a) -- ++(-0.5,-0.5) -- node[above] {$x:= x + i$} ++(-3.5,0) -- (l1);
   \draw[rounded corners] 
        (e11a) -- ++(0,-1) -- node[above] {$x:= x - i$} (l1);

   \draw[rounded corners] 
 	    (l1) -- ++ (1, -1) -- node[below] {\small $x > 42 \land $ $ i \neq 0 \land$ $y > 32$} (e11b);
   \draw[rounded corners] 
        (e11b) -- ++(-0.3,0.3) -- node[above] {$x:= x + i$} ++(-3.5,0) -- (l1);

\end{tikzpicture}}
  \end{center}
  \vspace{-3mm}
  \caption{Reactive program game structure with two system-controlled variables $x$ and $y$ and input variable $i$. We use  $\mathsf{skip}$ to denote the update $x := x; y := y$.   If the update to a variable is missing from the label of an update edge, its value remains unchanged.  Double-circles denote the B\"uchi accepting locations $\{l_2,l_3\}$.}
\label{fig:ex-buchi}
\end{figure}

\begin{example}\label{ex:buchi-motivating}
In the reactive program game depicted in Figure~\ref{fig:ex-buchi}, the Büchi specification defined by the set of locations $\{l_2,l_3\}$ requires the system to visit some of these two locations infinitely often.
Looping between $l_0$ and $l_3$ is possible as long as $y > 16$ at location $l_0$. 
Otherwise,  upon reaching $l_3$ with $y < 16$,  the program will have to transition to $l_4$ and revisiting either of $l_3$ or $l_2$ is not possible from this point on. 
From location $l_0$ the system can transition to $l_1$, from where,  similarly as in the game from Example~\ref{ex:intro}, the system can ensure reaching $l_2$ as long as $x \leq 42$ or $y \leq 32$.  
If $y > 32$ and $x >42$ in location $l_1$, the environment can prevent the system from decreasing $x$ by always setting $i$ to a positive value. 
Thus, if every time in location $l_0$ the system transitions to $l_1$ when $y \leq 32$ and to $l_3$ when $y > 32$  it can ensure that the set of locations $\{l_2,l_3\}$ is visited infinitely often and thus satisfy the given Büchi specification.
Our procedure successfully computes that every state of the form $(l_0,\assignment)$ is winning for the system.  As part of the computation,  attractor acceleration is applied to establish that from every state of the form $(l_0,\assignment)$ where  $\assignment(y)\leq 32$ the system has a strategy to enforce reaching $l_2$.  Note the presence of an unbounded strategy loop,  due to which the standard symbolic method for solving Büchi games is bound  to diverge.
\qed
\end{example}

In the above example, applying acceleration to  the attractor computation performed as part of the procedure for solving B\"uchi games,  suffices to ensure convergence.  In Section~\ref{sec:accel-buechi} we give an example where this is not enough, and describe an acceleration method for B\"uchi conditions.

\section{Technical Preliminaries}\label{sec:prelim}
\paragraph{Functions, Terms, and First-Order Logic.}

Let $\values$ be the set of all values of arbitrary types, 
$\functions := \{ f:\values^n \to \values ~|~ n \in \Nat \}$ be the set of all functions, and 
$\predicates := \{ p \in \functions \mid \mathit{Range}(p) \subseteq \Bool \}$ be the set of all predicates.
Let $\vars$ be the countably infinite set of all variables.
For $V \subseteq \vars$, we denote with $V':=\{x'\mid x \in V\}$ the set of primed variables such that $V \cap V' = \emptyset$.
Terms are used to describe functions and predicates. 
Let $\funcSymbols$ be the set of all function symbols and 
$\predSymbols \subset \funcSymbols$ the set of all predicate symbols. 
Function terms $\funcTerms$ are defined by the grammar 
$\funcTerms \ni \tau_f ::= f(\tau_f^1, \dots \tau_f^n) \:|\: x$ for $f \in \funcSymbols$ and $x \in \vars$, and we denote with $\predTerms$ the function terms of Boolean type.


A function $\assignment: \vars \to \values$ is called a \emph{variable assignment} (or simply \emph{assignment}).
The set of all assignments over a set of variables $V \subseteq \vars$ is denoted as $\assignments{V}$.
We denote the combination of two assignments $\assignment', \assignment''$ over  disjoint sets of variables by $\assignment' \uplus \assignment''$.
Given an assignment function $\assignment$,  a variable $x \in \vars$ and a value $v \in \values$,  we define the assignment function $\assignment[x := v]$ such that $\assignment[x := v](x) = v$ and $\assignment[x := v](y) = \assignment(y)$ for all $y \in \vars$ with $y \neq x$.

A function $\interpretation: \funcSymbols  \to \functions$ is called an \emph{interpretation of the function symbols} (or simply \emph{interpretation}).
We require 
 $\interpretation(p) \in \predicates$ for $p \in \predSymbols$.
The set of all interpretations of a set of function symbols $\Sigma \subseteq \funcSymbols$ is denoted as $\interpretations{\Sigma}$.
We denote the combination of two interpretations $\interpretation', \interpretation''$ over  disjoint symbol sets by $\interpretation' \uplus \interpretation''$.
The evaluation of function terms $\eval{\assignment,\interpretation}: \funcTerms \to \values$ is defined by $\eval{\assignment,\interpretation}(x) := \assignment(x)$ for $x \in \vars$, $\eval{\assignment,\interpretation}(f(\tau_0, \dots \tau_n)) := \interpretation(f)(\eval{\assignment,\interpretation}(\tau_0), \dots \eval{\assignment,\interpretation}(\tau_n))$ for $f \in \funcSymbols$. 

We denote the set of all first-order formulas as $\FOLX$.
Let $\varphi$ be a formula and $X = \{x_1,\ldots,x_n\} \subseteq \vars$ be a set of variables.
For a quantifer $Q \in \{\exists, \forall\}$, we write $Q X.\varphi$ as a short-cut for $Q x_1.\ldots Q x_n.\varphi$.  
We denote by $\QFX$ the set of all quantifier-free formulas in $\FOLX$.
We write $\varphi(X)$ to denote that the free variables of $\varphi$ are a subset of $X$.  
We also denote with $\FOL{X}$ and $\QF{X}$ the set of formulas (respectively quantifier-free formulas) whose free variables belong to $X$.
Given variables $x_1,\ldots,x_n \in \vars$, constant function terms with arity zero $c_1, \ldots,c_m \in \funcTerms$ and function terms $\tau_f^1,\ldots,\tau_f^{n+m}, \in \funcTerms$, we denote with $\varphi[x_1 \mapsto \tau_f^1,\ldots, x_n \mapsto \tau_f^n, c_1 \mapsto \tau_f^{n+1}, c_m \mapsto \tau_f^{n+m}]$ the formula obtained from $\varphi$  by the simultaneous replacement of all free occurrences of each $x_i$ with the respective term $\tau_f^i$ and of each occurrence $c_i$ by $\tau_f^{n+i}$.  
We denote with $\FOLentails: \assignments{\vars} \times \interpretations{\funcSymbols} \times \FOLX$ the entailment of first-order logic formulas.
A \emph{first-order theory} $T \subseteq \interpretations{\funcSymbols}$ restricts the possible interpretations of function and predicate symbols.
Given a theory $T$, for a formula $\varphi(X)$ and assignment $\assignment \in \assignments{X}$  we define that  
$\assignment \FOLentailsT{T} \varphi$ if and only if $\assignment, \interpretation \FOLentails \varphi$ for all 
$\interpretation \in T$.

\paragraph{Two-Player Games of Infinite Duration.}
We consider two-player games between a \emph{system player} and an \emph{environment player}.
A  \emph{game structure} is a tuple $G = (\states,  M_\e, M_\s, \rho)$, 
where $\states$ is a set of states,  
$M_\e$ and $M_\s$ are the sets of possible moves for Player~\env\ and Player~\sys\ respectively,  and 
$\rho \subseteq \states \times M_\e \times M_\s \times \states$ is a \emph{transition relation}
where:
(1) for all $s \in \states$, there exist $m_\e \in M_\e$,  $m_\s \in M_\s$  and $s' \in \states$ such that $(s,m_\e,m_\s,s') \in \rho$,  and 
(2) for all $s \in \states$,  $m_\e \in M_\e$ and $m_\s \in M_\s$,  if $(s,m_\e,m_\s,s_1) \in \rho$ and $(s,m_\e,m_\s,s_2) \in \rho$ then $s_1 = s_2$.
Condition (1) states that every state  has a successor, and  (2) states that the moves chosen by the two players uniquely determine a successor.
We define the functions $\enabled{\e} : \states \to \power{M_\e}$ and $\enabled{\s} : \states \times M_\e \to \power{M_\s}$ that indicate the enabled moves of a player:
$\begin{array}{lll}
  \enabled{\e}(s) & := & \{m_\e \in M_\e \mid \exists m_\s\in M_\s. \exists s' \in \states. \;(s,m_\e,m_\s,s')\in \rho\}, \\
  \enabled{\s}(s,m_\e) & := & \{m_\s \in M_\s \mid \exists s' \in \states. \;(s,m_\e,m_\s,s')\in \rho\}.
\end{array}$
%

A game on $G$ is played by Player~\env\  and Player~\sys\  as follows.
In a state $s\in \states$,  
Player~\env\ chooses a move $m_\e \in \enabled{\e}(s)$,  
Player~\sys\ chooses a move $m_\s \in \enabled{\s}(s,m_\e)$.
These choices define the next state $s'$ such that $(s,m_\e,m_\s,s')\in \rho$,   
The game then continues from $s'$.
The resulting infinite sequence $\pi = s_0,s_1,s_2,\ldots \in \states^\omega$ of states
is called a \emph{play}.
For $p \in \{\env,\sys\}$ we define $1 - p := \sys$ when $p = \env$,  and $1- p := \env$ when $p = \sys$.
%
A \emph{strategy for Player~\env}\ is a function 
$\sigma_\e: \states^+ \to M_\e$ where $\sigma_\e(s_0,\ldots,s_n) = m_\e$ implies $m_\e \in \enabled{\e}(s_n)$.  
A \emph{strategy for Player~\sys}\ is a function 
$\sigma_\s: \states^+ \times M_\e \to M_\s$ such that $\sigma_\s((s_0,s_1,\ldots,s_n),m_\e) = m_\s$ implies $m_\s \in \enabled{\s}(s_n,m_\e)$.  
We denote with $\str{p}{G}$ the set of all strategies for Player $p \in \{\env,\sys\}$.

Given $s \in \states$ and strategies $\sigma_\e$ and $\sigma_\s$  for the two players,  we denote with $\outcome{s}{\sigma_\e}{\sigma_\s}$ the unique play $s_0,s_1,s_2,\ldots$ such that $s_0 = s$, and for all $i \in \mathbb N$ there exist $m_\e \in M_\e$ and $m_\s \in M_\s$ such that $\sigma_\e(s_0,s_1\ldots,s_i) = m_\e$,   $\sigma_\s((s_0,s_1\ldots,s_i),m_\e) = m_\s$ and $(s_i,m_\e,m_\s,s_{i+1}) \in \rho$.  
Given a $\sigma_p \in \str{p}{G}$ and $s\in \states$, we define $\plays(s,\sigma_p) := \{\pi \in \states^\omega \mid \exists \sigma_{1-p} \in \str{1-p}{G}.\outcome{s}{\sigma_\env}{\sigma_{\sys}} = \pi\}$. 
%
An \emph{objective} is a set $\Omega \subseteq \states^\omega$. 
The set of \emph{states winning for Player $p$} with respect to objective $\Omega$ is 
$W_p(G,\Omega):= \{s \in \states \mid \exists \sigma \in \str{p}{G}. \plays(s,\sigma) \subseteq \Omega\}.$

\section{Reactive Program Games}\label{sec:games}
In this section, we define \emph{reactive program games}, a symbolic model that we use to  specify and synthesize reactive programs that operate over infinite domains.
 Figure~\ref{fig:ex-motivating},  Figure~\ref{fig:ex-combined},  Figure~\ref{fig:ex-buchi},  and  Figure~\ref{fig:ex-rpg}  depict examples of reactive program games.
The interpretation of the function symbols appearing in a reactive program game is required to conform to a given first-order theory $T$. 
Unless stated otherwise, we consider the theories of linear (integer or real) arithmetic.
The reactive program game in Figure~\ref{fig:ex-motivating},  is defined in the theory of linear integer arithmetic. 

\begin{definition}[Reactive Program Game Structure]\label{def:reactive-program-games}
A \emph{reactive program game structure} is a tuple $\mathcal G = (T,\inputs, \cells, L, \Inv,\delta)$ with the following components.
$T$ is a first-order theory.
$\inputs \subseteq \vars$ is a finite set of \emph{input variables}.
$\cells \subseteq \vars$ is a finite set of \emph{program variables} where $\inputs \cap \cells  = \emptyset$.         
$L$ is a finite set of \emph{game locations}.
$\Inv: L \to \FOL{\cells}$ maps each location to a \emph{location invariant}.
$\delta \subseteq L \times \QF{\cells \cup \inputs} \times (\cells \to \funcTerms) \times L$ is a finite \emph{symbolic transition relation}, where for every  $l \in L$ the set of \emph{outgoing transition guards} 
$\guards(l) := \{ g \mid \exists u, l'.~(l, g, u , l') \in \delta \}$ satisfies the conditions:
\begin{itemize}
    \item[(1)] $\bigvee_{g \in \guards(l)} g \equiv_{T} \top$,  and for all $g_1, g_2 \in \guards(l)$ with $g_1 \neq g_2$ it holds that  $g_1 \land g_2 \equiv_{T} \bot$,
    \item[(2)] for all $g, u, l_1, l_2$,  if $(l, g, u, l_1) \in \delta$ and $(l, g, u, l_2) \in \delta$, then  $l_1 = l_2$, and
    \item[(3)] for every $l \in L$ and $\assmt{x} \in \assignments{\cells}$ such that $\assmt{x} \FOLentailsT{T} \Inv(l)$ there exist a transition $(l, g, u, l') \in \delta$ and  $\assmt{i} \in \assignments{\inputs}$ such that $\assmt{x} \uplus \assmt{i}\FOLentailsT{T} g$ and  $\assmt{x'} \FOLentailsT{T} \Inv(l')$ where $\assmt{x}'(x) =  \eval{\assmt{x}\uplus\assmt{i}}(u(x))$ for every $x \in \cells$.
\end{itemize}
\end{definition}
The requirements on  $\delta$ imply for each $l\in L$ that: 
(1) the guards in $\guards(l)$ partition the set $\assignments{\cells \cup \inputs}$.
(2) each pair of $g \in \guards(l)$ and update $u$  can label at most one outgoing transition from $l$, and
(3) if there is an assignment satisfying the location invariant at $l$, then there is an input assignment for which there is a possible transition,  i.e., there are no dead-end states.
We define
\begin{itemize}
\item the set of \emph{input assignments} in $\mathcal G$ as $\inputa_{\mathcal G} := \assignments{\inputs}$, and
\item the set of \emph{updates} in $\mathcal G$ as $\updates_{\mathcal G} := \{u \in \cells \to \funcTerms \mid \exists l,g,l'.(l, g, u, l') \in \delta\}$.
\end{itemize}

The semantics of the reactive program game structure $\mathcal G$ is a possibly infinite-state game structure.
The set of \emph{states of the reactive program game structure} $\mathcal G$ consists of pairs $(l,\assmt{x})$ of game location $l$ and assignment $\assmt{x}$ to the program variables $\cells$.
The moves of Player~$\env$ (modeling the environment) are the input assignments $\inputa_{\mathcal G}$ and are potentially infinitely many.
Player~$\sys$ (corresponding to the program being synthesized) chooses the updates to the program variables (from the set $\updates_{\mathcal G}$) and has, therefore, by definition only finitely many possible moves. 

{\it Remark:} Location invariants are used to define the state-space of a reactive program game, and thus they provide modeling flexibility. They also enable the symbolic construction of sub-games, which are useful in game-solving procedures.
When all the location invariants in a game structure are $\top$, as,  for instance, in our examples,  we omit them for the sake of brevity.

\begin{definition}[Semantics of Reactive Program Game Structures]\label{def:rpgs-semantics}
Let 
$\mathcal G = (T,\inputs, \cells, L, \Inv, \delta)$ be a reactive program game structure.
The semantics of $\mathcal G$ is the game structure 
$\sema{\mathcal G} = (\states, M_\e, M_\s, \rho)$ where\looseness=-1
    \begin{itemize}
        \item   $\states := \{ (l,\assmt{x}) \in L \times \assignments{\cells} \mid \assmt{x} \FOLentailsT{T} \mathit{Inv}(l)\}$;
        \item   $M_\e := \inputa_{\mathcal G}$, $M_\s := \updates_\mathcal{G}$;
        \item   the transition relation $\rho$ contains $((l,\assmt{x}), \assmt{i}, u,(l',\assmt{x}'))$ if and only if 
           \begin{itemize}
        	\item there exists $g \in \guards(l)$ such that $(l,g,u,l') \in \delta$ and $\assmt{x}\uplus\assmt{i} \FOLentailsT{T} g$, and
        	\item $\assmt{x}'(x) =  \eval{\assmt{x}\uplus\assmt{i}}(u(x))$ for every $x \in \cells$,  and  $\assmt{x}' \FOLentailsT{T} \mathit{Inv}(l')$.
        \end{itemize}
    \end{itemize}
\end{definition}

The transition relation $\rho$ is well-defined as the conditions on $\delta$ in Definition~\ref{def:reactive-program-games}  ensure that  given $(l,\assmt{x}) \in S$, $\assmt{i} \in \inputa_{\mathcal G}$, and $u \in \updates_{\mathcal G}$, the successor location $l'$ and program variable assignment $\assmt{x}'$ are uniquely determined.
The successor assignment $\assmt{x}'$  is obtained by updating the value of each program variable $x \in \cells$ according to the corresponding update term $u(x)$ and assignment $\assmt{i}$ to the input variables.
For a state $s= (l,\assmt{x}) \in \states$, we denote with $\loc(s):=l$ the location of $s$.

We consider the realizability problem for reactive program games, formally defined below.

\begin{tcolorbox}[
  colback=gray!5!white,
  colframe=gray!75!black,
  title={Realizability and Program Synthesis for Reactive Program Games}]
Given a reactive program game structure $\mathcal G = (T,\inputs, \cells, L, \Inv, \delta)$,  objective $\Omega$ for   Player~\sys,  and $l \in L$,  the \textbf{\emph{realizability problem}} is to determine if $(l,\assmt{x}) \in W_{\sys}(\sema{\mathcal G},\Omega)$ for every $\assmt{x} \in \assignments{\cells}$ with $\assmt{x} \FOLentailsT{T}\mathit{Inv}(l)$.
The \textbf{\emph{program synthesis problem}} is to compute a strategy for Player~\sys\ that is winning from every $(l,\assmt{x}) \in W_{\sys}(\sema{\mathcal G},\Omega)$. 
\end{tcolorbox}

\section{Symbolic Procedures for Reactive Program Games}\label{sec:solving}
Given a reactive program game structure $\mathcal G = (T,\inputs, \cells, L, \Inv, \delta)$ and an objective $\Omega$ for Player $p$,  the \textbf{\emph{game solving problem}} is to compute the set of states $W_p(\sema{\mathcal G},\Omega)$.
The realizability question for given location $l$ can be answered by checking if this set contains all states $s \in \states$ with $\loc(s) = l$.

We present procedures for solving reactive program games with important types of objectives considered in reactive synthesis.  Similarly to the respective algorithms for finite-state games, a building block of these procedures is the computation of the so-called attractor sets.  
We lift the classical algorithms to infinite-state games by employing a symbolic attractor computation procedure.  Since the corresponding game-solving problems are generally undecidable, the game-solving procedures are not guaranteed to terminate.  In the next section, we propose an acceleration technique, which,  when it succeeds, enforces convergence.

\subsection{Symbolic Representation and Operations}

We now present the basic building blocks of our procedures for solving reactive program games:  the symbolic representation of sets of states and the necessary operations on this  representation.

We represent and manipulate possibly infinite sets of states symbolically,  
using formulas in \FOL{\cells} to represent sets of assignments to the variables in $\cells$.  
We define our \emph{symbolic domain}  $\symstates := L \to \FOL{\cells}$ to be the set of functions mapping locations to first-order formulas in \FOL{\cells}. 
The semantics $\sema{\cdot}: \symstates \to \power{\states}$ of an element $d$ of $\symstates$ is defined by
   $\sema{d} := \{ (l, \assmt{x}) \in \states \mid \assmt{x} \FOLentailsT{T}  d(l) \land \assmt{x}\FOLentailsT{T}  \Inv(l)\}.$
Note that locations are treated explicitly. 

We perform set operations on $\symstates$ symbolically, per location.
Formally,  for $d_1,d_2 \in \symstates$ we define 
$\neg d_1 := \lambda l.\;\neg d_1(l)$,  
$d_1 \wedge d_2 := \lambda l.\;d_1(l) \wedge d_2(l)$,  
$d_1 \vee d_2 := \lambda l.\;d_1(l) \vee d_2(l)$, and write 
$d_1 \equiv_{T} d_2$ iff $\sema{d_1} = \sema{d_2}$.
We define $d[l \mapsto \varphi] := (\lambda l'.~\text{if}~l = l'~\text{then}~\varphi~\text{else}~d(l'))$ for $d \in \symstates$, which is obtained from $d$ by changing the symbolic value at a given location.
We write $\{ l_1 \mapsto \varphi_1, \dots, l_n \mapsto \varphi_n \}$
for the element $d \in \symstates$ where $d(l_i) = \varphi_i$, for $i \in \{1, \dots, n\}$ and $d(l) = \bot$ for $l \not\in \{l_1, \dots, l_n\}$.

\paragraph{Attractor.} 
Let  $R \subseteq \states$ be a set of states.  
The set of states from which  Player~$p$ can enforce reaching a state in 
$R$ is called the \emph{Player~$p$-attractor for $R$ in $\sema{\mathcal G}$} 
and is denoted by
$ \mathit{Attr}_{\sema{\mathcal G},p}(R)$.  Formally, 
\[ \mathit{Attr}_{\sema{\mathcal G},p}(R) := \{s \in \states \mid \exists \sigma \in \str{p}{\sema{\mathcal G}}.\forall \pi \in \plays(s,\sigma).\exists n\in \Nat.\;\pi[n]\in R\}.\]

In the finite-state case,  attractors are computed by a fixpoint iteration using the so-called \emph{controllable predecessor operators}.  
We define their symbolic counterparts
\[ \cpre{\mathcal G,\env}{d} := \lambda l.~ \Inv(l) \land \exists\inputs.~\bigwedge_{(l, g, u, l') \in \delta} \left( g \land \Inv(l')[u] \to d(l')[u] \right) \]
\[ \cpre{\mathcal G,\sys}{d} := \lambda l.~ \Inv(l) \land \forall\inputs.~\bigvee_{(l, g, u, l') \in \delta} \left( g \land \Inv(l')[u] \land d(l')[u] \right) \]
for $d \in \symstates$ where $\varphi[u] := \varphi[c \mapsto u(c) \mid c \in \cells]$ for $u  \in \updates_\mathcal{G}$ applies the substitution defined by $u \in \cells \to \funcTerms$ to $\varphi \in \FOL{\cells}$ resulting in the formula obtained by the simultaneous replacement of all $x \in \cells$ with the respective term $u(x)$.
By definition,  we have that $\mathit{CPre}_{\mathcal G,\env},\mathit{CPre}_{\mathcal G,\sys}: \symstates \to \symstates$.

\begin{figure}[t]
\begin{algorithm}[H]
\SetAlgoVlined
    \SetKwProg{Fn}{function}{}{}
    \DontPrintSemicolon
    \Fn{\textsc{Attractor}(
		 $\mathcal G = (T,\inputs, \cells, L, \Inv,\delta)$,  
    	 $\mathit{p} \in \{\sys,\env\}$,  
		 $d \in \symstates$)}{
		\setcounter{AlgoLine}{0}
		\nl $a^0$ := $\lambda l.~\bot$;
        $a^1$ := $d$\;
        \nl \For{$n=1,2,\ldots$}{
			\nl	\lIf{$a^{n} \equiv_{T} a^{n-1}$}{\Return $a^n$}
			\nl $a^{n+1} := a^n \lor \cpre{\mathcal G,p}{a^n}$} 		
	}
\caption{Symbolic semi-algorithm for computing $\mathit{Attr}_{\sema{\mathcal G},p}(\sema{d})$.}\label{algo:attractor}	
\end{algorithm}
\vspace{-4mm}
\end{figure}

\begin{figure}[b]
\vspace{-3mm}
\begin{subfigure}[b]{0.45\textwidth}
\centering
\scalebox{0.8}{%
\begin{tikzpicture}[->,>=stealth',shorten >=1pt,auto,node distance=3cm,scale=0.9]
\tikzstyle{every state}=[fill=none,draw=black,text=black,inner sep=1.5pt, minimum size=11pt,thick]
  \node[state] (l0) at (0,0) {$l_0$};
  \node[state] (l1) at (3.5,0) {$l_1$};
  
  \node[fill=black,draw=black,minimum size=0.5pt] (e00) at (0,1.5) {};
  \node[fill=black,draw=black,minimum size=0.5pt] (e01) at (2,.5) {};
  \node[fill=black,draw=black,minimum size=0.5pt] (e10) at (1.2,0) {};
  \node[fill=black,draw=black,minimum size=0.5pt] (e11) at (5.5,0) {};

  \path
    (l0) edge[bend left=25] node[above,midway,sloped] 
        {\small $i \geq 5$} 
    (e00)    
    (e00) edge node[above,midway,sloped] 
        {\small $\mathit{skip}$} 
    (l0)        
    (l0) edge[bend left=15] node[above, align=center] 
        {\small $i < 5$} 
    (e01)
    (e01) edge[bend left=15] node[above, align=center] 
        {\small $x := i$} 
    (l1)
    (l1) edge node[below, midway,align=left] 
        {\small $x \not\in [3, 5]$} 
    (e10)
    (e10) edge node[below, midway,align=left] 
        {\small $\mathit{skip}$} 
    (l0)    
    (l1) edge node[above, midway,sloped] 
    	{\small $ x \in [3, 5]$}
    (e11)         
  ;
      \draw[rounded corners] 
        (e11) -- ++(0,0.7) -- node[above] {\small $x:=0.5 \cdot x$} ++(-2,0) -- (l1);
      \draw[rounded corners] 
        (e11) -- ++(0,-.5) -- node[below] {\small $\mathit{skip}$} ++(-2,0) -- (l1);

\end{tikzpicture}}
\vspace{-2mm}			
\caption{The game structure.}\label{fig:ex-rpg}
\end{subfigure}
\hfill
\begin{subfigure}[b]{0.54\textwidth}
\centering
\scalebox{1.0}{%
\begin{tabular}[b]{l} 
$a_0(l_1) = a_1(l_1) = \bot$\\
$a_2(l_1) = x \not \in[3,5]$\\
$a_3(l_1) = x \not \in[3,5] \lor x \in [3,5] \land 0.5 \cdot x \not\in [3,5]$\\
$\quad\equiv x \not \in[3,5] \lor x \in [3,5] \land x \not\in [6,10] \equiv \top$\\
$a_4(l_1) = \top $ (fixpoint as $\equiv a_3(l_1)$)
\end{tabular}}
\vspace{-2mm}			
\caption{Iterations of \textsc{Attractor} for location $l_1$.}
\label{fig:table-game-struct-2}
\end{subfigure}
\vspace{-6mm}
\caption{
(right) A reactive program game with locations $\{ l_0,l_1 \}$, input $i$ and program variable $x$, both of type Real. 
(left) The iterative computation of $\textsc{Attractor}(\mathcal{G}, \sys, \{l_0 \mapsto \top\})$ for location $l_1$.
}\label{fig:full-game-struct-2}
\end{figure}

Algorithm~\ref{algo:attractor} can be used to compute $ \mathit{Attr}_{\sema{\mathcal G},p}(R)$ symbolically, given a symbolic representation of $R$. 
 Figure~\ref{fig:table-game-struct-2} shows an example of such a computation, and the next proposition states its soundness.

\begin{proposition}
\label{prop:correctness-attractor}
Let $\mathcal G$ be a reactive program game structure, 
$\mathit{p} \in \{\sys,\env\}$ and  $d \in \symstates$.
If \textsc{Attractor}($\mathcal G$,$p$,$d$) terminates and returns $\mathit{attr}$,   then $\sema{attr} = \mathit{Attr}_{\sema{\mathcal G},p}(\sema{d})$.
\end{proposition}

\subsection{Symbolic Game Solving}
In reactive program games we consider \emph{objectives defined in terms of the locations appearing in a play}.
Below we recall the definitions of common types of objectives and the classical algorithms for solving such games, formulated symbolically in the context of reactive program games.

\paragraph{Reachability and Safety Games.}
The \emph{reachability objective}  $\reach(R)$ 
for  $R \subseteq L$,  
requires that some state with location in $R$ is visited eventually.
Formally,  $
\reach(R) := \{\pi\in \states^\omega \mid \exists n\in \Nat.\; \loc(\pi[n])  \in R \}$. 
The dual, \emph{safety objective} $\safe(S)$
for $S \subseteq L$,  
 requires that only locations in $S$ are visited by the play.
Formally,
$\safe(S) := \{\pi\in \states^\omega \mid \forall n\in \Nat.\; \loc(\pi[n]) \in S \}$. 

Reactive program games with reachability objectives for a Player~$p$ can be solved by applying Algorithm~\ref{algo:attractor} to compute the Player~$p$-attractor for the set of states with locations in $ R$. 
More concretely, Proposition~\ref{prop:correctness-attractor} directly implies that 
$W_p(\sema{\mathcal G},\reach(R))= \sema{\textsc{Attractor}(\mathcal{G},p, \{ l \mapsto \Inv (l) \mid l \in R \})}$ holds if the iteration in \textsc{Attractor}($\mathcal G$,$p$,$\{ l \mapsto \Inv (l) \mid l \in R \}$) terminates.

Since safety is dual to reachability,  we can solve a reactive program game with safety objective $\safe(S)$ for Player~$p$ by solving the game with a reachability objective for Player~$1-p$ where the set of goal locations is $L \setminus S$. More precisely,
we call Algorithm~\ref{algo:attractor} as
\textsc{Attractor}($\mathcal G$,$1-p$,($\{ l \mapsto \Inv(l) \mid l \not\in S \}$). 
If it terminates and returns $\mathit{attr}$, then we have that $
W_p(\sema{\mathcal G},\safe(S)) = \sema{\neg \mathit{attr} \wedge \Inv}.$

\paragraph{Büchi and Co-Büchi Games.}
The \emph{Büchi objective} $\buchi(B)$ 
for a set of accepting locations $B \subseteq L$,  
requires that $B$ is visited infinitely often:
$\buchi(B) := \{\pi\in \states^\omega \mid \forall m\in \Nat.\exists n > m.\; \loc(\pi[n]) \in B \}$. 
The dual, \emph{co-Büchi objective} $\cobuchi(C)$ for rejecting locations $C \subseteq L$ requires that $C$ is visited only finitely many times:
$\cobuchi(C) := \{\pi\in \states^\omega \mid \exists m\in \Nat.\forall n \geq m.\; \loc(\pi[n]) \not\in C \}$.

Employing the symbolic attractor computation procedure in Algorithm~\ref{algo:attractor}, we lift the classical algorithm for solving finite-state games with Büchi and co-Büchi objectives to a procedure for solving reactive program games with these objectives.  
The procedure, given in  Algorithm~\ref{algo:buchi}, is based on a nested fixpoint computation, as the classical algorithm.
The inner fixpoint iteration computes attractor sets for the Player $p$ with Büchi objective $\buchi(B)$, and the outer fixpoint iteration computes increasing underapproximations of the set of winning states for the Player $1-p$ with co-Büchi objective $\cobuchi(B)$. 
At each iteration,  we first compute $a^n$, which represents the states from which Player $p$ can enforce a visit to $f^n$,  by calling Algorithm~\ref{algo:attractor}. 
Then,  we compute $w^{n}_{1- p}$, which represents the states from which Player~$1-p$ can prevent Player~$p$ from \emph{revisiting} $f^n$.  This is done by calling Algorithm~\ref{algo:attractor} to compute an attractor for Player~$1-p$.

\begin{figure}[t!]
\begin{algorithm}[H]
\SetAlgoVlined
    \SetKwProg{Fn}{function}{}{}
    \DontPrintSemicolon
    \Fn{\textsc{SolveB\"uchiGame}(
		 $\mathcal G = (T,\inputs, \cells, L, \Inv,\delta)$,  
    	 $\mathit{p} \in \{\sys,\env\}$,  
		 $f \in \symstates$)}{
		\setcounter{AlgoLine}{0}
		\nl $f^0$ := $\lambda l.~\top$;
	    $f^1 $ := $f$;
	     $w_{1-p}^0  :=  \lambda l.  \bot$\;
	     \nl \For{$n=1,2,\ldots$}{
	     \nl	\lIf{$f^{n} \equiv_{T} f^{n-1}$}{\Return $w_{1-p}^{n-1}$}
	     \nl $a^n := \textsc{Attractor}(\mathcal G, p, f^n)$\;
	     \nl $w^{n}_{1- p} := \textsc{Attractor}(\mathcal G,1- p,\neg  \cpre{\mathcal G,p}{a^n})$\;
	     \nl $f^{n+1} :=  f^n \wedge (\Inv \land \neg w^{n}_{1- p})$ 
	     }

	}
		 \caption{Symbolic semi-algorithm for computing the set of states winning for Player~$1-p$ in a game with a B\"uchi objective for Player~$p$ with set of accepting states represented by $f$.}
\label{algo:buchi}
\end{algorithm}
\vspace{-5mm}
\end{figure}

To solve a reactive program B\"uchi game defined by a set of locations $B$ in $\mathcal G$,  we execute Algorithm~\ref{algo:buchi} with $f:= \{ l \mapsto \Inv(l) \mid l \in B \}$.
If the computation reaches a fixpoint, that is,  for some iteration $m$ we have that $f^{m} \equiv_{T} f^{m-1}$ (and hence,  $w_{1-p}^{m} = w^{m-1}_{1-p}$),  then we have that $\sema{w^{m}_{1-p}} = W_{1-p}(\sema{\mathcal G},\cobuchi(B))$ and
 $\sema{\neg w^{m}_{1-p} \wedge \Inv} = W_{p}(\sema{\mathcal G},\buchi(B))$.
This follows from the correctness of the classical algorithm for solving Büchi games~\cite{2001automata,BloemCJ18}.

\subsection{Strategy Extraction}\label{sec:strategy-extraction}
When the answer to the realizability question for a reactive program game is positive, that is, 
for the given location $l_{\mathit{init}}$ all states $s \in \states$ with $\loc(s) = l_{\mathit{init}}$  are winning for Player~\sys, we want to extract a winning strategy for Player~\sys\ in the form of a program.
To this end, we lift the strategy-generation extensions of the classical algorithms on which the symbolic  procedures are based. \looseness=-1

We represent winning strategies for Player~\sys\ as simple GOTO programs,  with labels corresponding to locations,  and which contain \textcolor{purple}{\tt goto} statements,  \textcolor{purple}{\tt read} statements for the input variables, conditionals, and the selected updates from the game in form of (parallel) assignments to the program variables. 
Later on,  we will extend these programs with additional labels, auxiliary variables, and simple assignments to the auxiliary variables.
We now describe how the symbolic game-solving procedures described earlier in this section are extended to produce such programs.

The basic building block for strategy generation is the extraction of a program statement in a location $l \in L$ from a strategic decision based on the controllable predecessor $\mathit{CPre}_{\mathcal G,\sys}(d)(l)$ for some $d \in \symstates$. This requires a strategy that enforces reaching $d$  in one step from $l$.
Such a strategy selects a transition $(l, g, u, l')$ where the guard $g$ holds, and after the update $u$,  $d(l')$ and $\mathit{Inv}(l')$ hold. 
For example, for $d = \{ l' \mapsto x > 0 \}$, guard $x < 5$,  update $x := x + 1$,  and $\mathit{Inv}(l') = \top$,  we extract the program  statement $\texttt{if}~(x < 5 \land x + 1 > 0)~\texttt{then}~x := x + 1;~\texttt{\textcolor{purple}{goto}}~l'~\texttt{else} \dots$ which could then be followed by other such transition statements.  Note that before branching according to the choice of the strategy at a given location, we have to add \textcolor{purple}{\tt read} statements for the input variables.

For games with reachability objectives,  a winning strategy can be generated based on the attractor computation augmented with keeping a record of the ``layers'' of the attractor, that is the individual $a^1,a^2,\ldots$ until termination.  For each location $l\in L$ and states in $a^{n+1}(l) \land \lnot a^n(l)$,  that is, states added to the attractor in step $n+1$,  the winning strategy will select a transition based on the controllable predecessor $\mathit{CPre}_{\mathcal G,\sys}( a^n)(l)$,  resulting in a program statement as described above.

For safety games, we first compute the set of states $W_{\sys}(\sema{\mathcal G},\safe(S))$ winning for Player~\sys, symbolically represented by some $w \in \symstates$.  Then,  for each location $l \in L$,  we generate a program statement representing a strategy to remain in $w$, that is, based on $\mathit{CPre}_{\mathcal G,\sys}(w)(l)$.

The procedure for solving games with Büchi and co-Büchi objectives can also be extended with the necessary bookkeeping to generate winning strategies by using strategies (programs) generated from the controllable predecessor operator and attractor computation as building blocks.

\paragraph{What about the environment?}
Winning strategies for Player~\env\ are less useful as we are usually interested in strategies that give us an implementation of the desired system.  They can be helpful as counterexamples, but for reactive program games,  extracting a strategy for Player~\env\ is more difficult,  as the set of moves   $M_\e := \inputa_{\mathcal G}$ is potentially infinite.  

\section{Enforcement Acceleration}\label{sec:acceleration}

The symbolic procedure for attractor computation,  and hence the procedures for solving games, presented in the previous section do not always terminate when the set of states $\states$ is infinite.

As we explained in Section~\ref{sec:examples},  the attractor computation for the  reactive program game structure in Figure~\ref{fig:ex-motivating} and the reachability winning condition for Player \sys\  to reach $l_G$, does not terminate. 
Applying the method from the previous section as $\textsc{Attractor}(\mathcal{G}, \sys,\{l_G \mapsto \top\})$, we get:
$a^2 = \{ l_0 \mapsto (x \leq 42), l_G \mapsto \top \}$,
$a^3 = \{ l_0 \mapsto (\forall i. x \leq 42 \lor i = 0 \lor x + i \leq 42 \lor x - i \leq 42), l_G \mapsto \top \}$, which simplifies to $\{ l_0 \mapsto (x \leq 43), l_G \mapsto \top\}$,  
$a^4  =\{l_0 \mapsto (x \leq 44), l_G \mapsto \top\}$, 
$a^5  =\{l_0 \mapsto (x \leq 45), l_G \mapsto \top\}$,\ldots. 
A fixpoint is never reached as $\sema{a^n}$ keeps growing in every iteration.

However, arguing for termination on an intuitive level is done quite simply as already shown in Section~\ref{sec:intro}.
If $x \leq 42$ or $i = 0$ in location $l_0$, then $\{l_G \mapsto \top \}$ is reached in one step.
Otherwise,  by choosing the correct transition in location $l_0$,  Player~\sys\ can force $x$ to decrease and go back to location $l_0$.
Since the game is back in $l_0$, decreasing $x$ can be iterated, eventually leading to $x \leq 42$.

While we cannot ensure termination in general, we will augment the attractor computation with an acceleration operation. 
This operation should extend the computed attractor set with states from infinitely many levels in finitely many steps.
Since the attractor captures strategic decisions, this acceleration operation allows us to handle unbounded strategy loops as the one in Figure~\ref{fig:ex-motivating}.

For the rest of this section, we fix a first-order logical theory $T$.

\subsection{Accelerating Symbolic Attractor Computation}\label{sec:acceleration-attractor}

From a more general perspective,  the termination argument used in our running example above states the following: \emph{If in some location $l$, starting at some state not part of the goal states, Player~$p$ can enforce coming back to $l$ and make progress towards the goal, then Player~$p$ can enforce reaching the goal eventually.}
Our aim is to employ such reasoning to accelerate the computation performed by Algorithm~\ref{algo:attractor}.  Suppose that $a^n$ is the symbolic representation of the current subset of $\mathit{Attr}_{\sema{\mathcal G},p}(\sema{d})$ at the $n$-th iteration of the loop in Algorithm~\ref{algo:attractor}.
Let $l \in L$ be a location in the reactive program game.  
The goal of the acceleration operation applied to location $l$ is to expand the set $\sema{a^n}$ by adding a set of states at location $l$ from which Player~$p$ can enforce reaching $\sema{a^n}$ by enforcing a loop at location $l$ to be executed some number of times.
To perform such reasoning automatically,  we first introduce the new notion of  \emph{acceleration lemmas} that allows us to formalize inductive arguments such as the one from our running example.  
Next, to perform the acceleration computation at a given location $l$, we construct from $\mathcal{G}$ a new game structure, called a \emph{loop game} which we use to reason about  the loops from location $l$ back to itself.  
In the following,  we first give the formal definitions of these concepts and then proceed with the description of the procedure for attractor acceleration.

\subsubsection*{Acceleration Lemmas}
An acceleration lemma, formally introduced below, is a triple $(\base, \step, \conc)$ of \FOLX\ formulas. 
It formalizes an inductive statement as follows.
The \emph{base condition} $\base$ characterizes a set of target states.
The \emph{step relation} $\step$ is a relation between states.
The \emph{conclusion} $\conc$ characterizes states from which every sequence of states in which each pair of consecutive states conform to the relation $\step$ will necessarily reach the target set $\base$.
Thus, intuitively,  the relation $\step$ captures a ranking argument for establishing the reachability of $\base$ starting from $\conc$.

\begin{definition}[Acceleration Lemma]\label{def:lemma}
An \emph{acceleration lemma} is a triple $(\base, \step, \conc)$  of formulas $\base \in \FOL{V}$,   $\step \in \FOL{V \cup V'}$,  and $\conc \in \FOL{V}$ for some $V \subseteq \vars$,  such that: 
\begin{itemize}
\item
For every infinite sequence
$\alpha \in \assignments{V}^\omega$, if it holds that  
$\alpha[0] \FOLentailsT{T} \conc$ and 
$\combine{\alpha[i]}{\alpha[i+1]} \FOLentailsT{T} \step$ for all $i \in \Nat$, 
then there exists some $k \in \Nat$ such that $\alpha[k] \FOLentailsT{T} \base$,

\item and, 
for every $\assignment \in \assignments{V}$ with $\assignment \FOLentailsT{T} \conc$ and $\assignment \not\FOLentailsT{T} \base$,  and 
for every $\assignment'\in \assignments{V}$ with $\combine{\assignment}{\assignment'} \FOLentailsT{T} \step$,  it holds that
$\assignment' \FOLentailsT{T} \conc$,
\end{itemize}
where 
$\combine{\assignment}{\assignment'}(x) := \assignment(x)$ for $x \in V$ and
$\combine{\assignment}{\assignment'}(x') := \assignment'(x)$ for $x' \in V'$.

We denote with  $\lemmas(V)$  the set of acceleration lemmas over the set of variables $V$.
\end{definition}

The first condition in the definition formalizes the intuition above. 
The second condition is necessary for ensuring that the step relation can be iterated. 
It requires that starting from any state that satisfies $\conc$ but not $\base$,  any successor with respect to the relation $\step$ is also in $\conc$.  

\begin{example}\label{ex:lemma}
Consider $V = \{x\}$ where $x$ is of integer type,  and let $\base := (x \leq 42)$,  $\step := (x' < x)$ and $\conc := \top$.  It is easy to see that $(\base, \step, \conc)$ is an acceleration lemma, since the relation $<$ is well-founded on $\{ z \in \mathbb{Z} \mid z \geq 42\}$.\qed
\end{example}

Acceleration lemmas express generic inductive statements in a logical form.
Employing an acceleration lemma  $(\base, \step, \conc)$ to accelerate the $n$-th iteration of the attractor computation at location $l$ will result in adding the states $\{(l,\assmt{x}) \in S \mid \assmt{x} \FOLentailsT{T} \conc\}$ to the computed attractor.
This operation is correct if the following conditions are satisfied.
The first condition is that \emph{$\base$ is already included in $a^n(l)$}.
For a given acceleration lemma this condition can be established by checking the validity of a logical implication.
The second condition requires that Player~$p$ can actually \emph{enforce the repeated iteration of the step relation} in  the game structure $\mathcal{G}$, against all possible behaviors of the opponent Player~$1-p$.   
More precisely,  this means that starting from any state $(l,\assmt{x})$ satisfying the formula $\conc$, Player~$p$ should have a strategy to reach a state $(l, \assmt{x}')$ such that $(\assmt{x},\assmt{x}')$ satisfies the relation $\step$, or reach the current subset $a^n$ of the attractor. 
Establishing  satisfaction of this condition is more difficult  because it amounts to establishing the existence of a local strategy for Player~$p$ to enforce the step relation.
We show how we can reduce this question to attractor computation in a modified game structure,  the so-called loop game at location $l$, defined next.


\subsubsection*{Loop  Games}
The \emph{loop game} at location $l$ is a reactive program game structure obtained from the reactive program game structure $\mathcal G$ and a given location $l$.  
The game structure $\loopgame(\mathcal G,  l, \locend)$ is constructed by adding a new  location $\locend \not\in L$ to 
$\mathcal G$ and redirecting all edges in $\mathcal G$ with target location $l$ to the new location $\locend$.  
Intuitively,  the loop game ``splits'' the loops from location $l$ to itself and captures all possible interactions between Player~$\sys$ and Player~$\env$ in $\mathcal G$ that start from location $l$.
The splitting of the loops allows us to reason about iterated behavior from location $l$ back to itself. 
More precisely,  establishing that Player~$p$ can enforce iterating the step relation in location $l$ in $\mathcal{G}$ reduces to establishing that in the loop game Player~$p$ can enforce reaching $\locend$ from $l$ while satisfying the step relation.
We do the latter by computing an attractor for Player~$p$ in the loop game on $\step$  from $\locend$ back to $l$.
We define loop games formally  as follows.

\begin{definition}[Loop Game]
Given a reactive program game structure $\mathcal G = (T,\inputs, \cells, L, \Inv,\delta)$,  a location $\locsplit \in L$, and a fresh location $\locend \not\in L$,  the \emph{loop game}  
is the reactive program game structure
$\loopgame(\mathcal G,  \locsplit, \locend) := (T,\inputs, \cells, L \cup \{\locend\}, \Inv',\delta')$, where 
 $\Inv': = \Inv \cup \{\locend \mapsto \Inv(\locsplit)\}$, and  
 $\delta' := \{ (l_1 , g, u, l_2) \in \delta ~\mid~ l_2 \neq \locsplit \} \cup \{ (l, g, u, \locend) ~\mid~ (l, g, u, \locsplit) \in \delta \}\cup \{ (\locend, \top, \lambda x.x, \locend)\}$. 
\end{definition}

\begin{example}\label{ex:lemma-loop}
 Figure~\ref{fig:ex-loop-game} depicts the loop game $\loopgame(\mathcal G,  l_0, \locend)$ of our running example in Figure~\ref{fig:ex-motivating}.
\end{example}


\subsubsection*{Attractor Acceleration with Given Acceleration Lemmas.}
We now give a high-level description of how the attractor computation for Player~$p$ is accelerated for a given location $l$ using a given acceleration lemma $(\base, \step, \conc)$.
We are applying acceleration at some step $n$ of the algorithm, and $a^n$ is the current subset of the attractor computed thus far.
To apply the acceleration lemma, we fist check that $\base$ is included in $a^n(l)$.
We then construct the loop game $\loopgame(\mathcal G,  l, \locend)$. 
In this loop game, we compute the attractor for Player~$p$ on the step relation $\step$ in $\locend$ (we  will see later how $\step$ is encoded as a state formula with the help of auxiliary variables). 
We then check that the attractor computed in the loop game includes the states with location $l$ that satisfy $\conc$.
If both conditions hold,  we know that starting at location $l$ in a state that satisfies $\conc$, 
Player~$p$ can enforce reaching  $a^n(l)$ by iterating the strategy (from the loop game) to enforce $\step$ from $l$ back to $l$.
To see this, note that by Definition~\ref{def:lemma} we know that this will eventually lead to reaching $\base$, and hence also $a^n$.
This justifies extending $a^n(l)$ with $\conc$,  that is,  setting $a^n(l) : = a^n(l) \vee \conc$. 
After that,  we continue with the attractor computation,  possibly applying acceleration again.
We now illustrate the use of acceleration lemmas  on a concrete example.

\begin{example}\label{ex:lemma-loop-app}
We use the method outlined above to accelerate the diverging attractor computation shown at the beginning of this section, at the point where $a^3 = \{ l_0 \mapsto (x \leq 43), l_G \mapsto \top\}$.

We want to apply the acceleration lemma from Example~\ref{ex:lemma} at location $l_0$. 
First, we check that the base condition $\base = (x \leq 42)$ of our acceleration lemma is a subset of $a^3(l_0) =  (x \leq 43)$, which is indeed the case.
Then,  we have to check that starting in $l_0$,  Player~\sys\ can either enforce the step relation and come back to $l_0$ or reach $a^3$. 
To do this, we consider the loop game $\loopgame(\mathcal G,  l_0, \locend)$.
The condition above is satisfied if Player~\sys\ can enforce reaching $\locend$ from $l_0$ in the loop game such that the relation $\step$  is satisfied by the initial (at $l_0$) and final (at $\locend$) values of the program variables  (or reach $a^3$).
Therefore, we compute the attractor for Player~\sys\ of $\{ l_0 \mapsto (x \leq 43), l_G \mapsto \top, \locend \mapsto x < e_x \}$ in the loop game.
Here $x < e_x$ corresponds to the step relation $\step = x' < x$, where $e_x$ represents the initial value of $x$ before the iteration. This ``shift'' is necessary as we are doing a backward computation. 
This attractor computation in the loop game yields $\{ l_0 \mapsto \forall i.~i \neq 0 \land x > 42 \to x + i < e_x \lor x - i < e_x, \dots \}$.
Since at $l_0$ we have completed the iteration, we can set $e_x := x$ and the formula simplifies to $\{l_0 \mapsto \top, \dots\}$.
This means that Player~\sys\ can indeed iterate enforcing the step relation or reach $a^3$.

The property of acceleration lemmas allows us to conclude that Player~\sys\ can reach $a^3$ starting from $\conc$ by iterating $\step$.
Hence,  we can add $\conc = \top$ to the attractor set at location $l_0$,  which results in reaching a fixpoint and the computation terminates.
\qed
\end{example}

\begin{figure}[b]
\vspace{-3mm}
\begin{center}
\begin{tikzpicture}[->,>=stealth',shorten >=1pt,auto,node distance=2.5cm]
  \tikzstyle{every state}=[fill=none,draw=black,text=black,inner sep=1.5pt, minimum size=16pt,thick,scale=0.8]
    \node[state] (l) at (0,0) {$l_0$};
    \node[state] (lE) at (-4.2,0) {$\locend$};
    \node[state] (lG) at (4.2, 0) {$l_G$};
    
   \node[fill=black,draw=black,minimum size=0.5pt] (eEE) at (-5.5,0) {};   
   \node[fill=black,draw=black,minimum size=0.5pt] (eGG) at (5.5,0) {};
   \node[fill=black,draw=black,minimum size=0.5pt] (e0G) at (2.5,0) {};
   \node[fill=black,draw=black,minimum size=0.5pt] (e0E) at (-2.5,0) {};
   
    \path (l) edge node[above,midway] 
              {\small $ x > 42 \land i \neq 0$} (e0E)
             (e0E) edge[bend right=10] node[above,] 
              {\small $ x := x + i$} (lE)           
             (e0E) edge[bend left=10] node[below] 
              {\small $ x := x - i$} (lE)
          (l) edge node[above] {\small $x \leq 42 \lor i = 0$} (e0G)
          (e0G) edge node[above] {\small $ x := x$} (lG)
          (lG) edge[bend left=10] node[above] {\small $\top$} (eGG)
          (eGG) edge[bend left=10] node[below] {\small $x := x$} (lG)
          (lE) edge[bend right=10] node[above] {\small $\top$} (eEE)
          (eEE) edge[bend right=10] node[below] {\small $x := x$} (lE);
\end{tikzpicture}
\end{center}
\vspace{-3mm}
\caption{Loop game  $\loopgame(\mathcal G,  l_0, \locend)$ for the game structure in Figure~\ref{fig:ex-motivating}.}
\label{fig:ex-loop-game}
\end{figure}

Thus far,  we illustrated the usefulness of acceleration lemmas in attractor computation. 
To transform this idea into a general procedure for attractor acceleration,  we have to account for the following aspects.
In general, the loop game can be complex itself, and the attractor computation in the loop game may also benefit from acceleration.  
Hence, our acceleration technique should support the \emph{nested applications of lemmas}.
Another crucial aspect of the procedure is the actual \emph{computation of acceleration lemmas}.
A simple way to realize this would be first to generate acceleration lemmas and then check their applicability with the described technique,  backtracking if the check is not successful.
However, this quickly becomes infeasible for large spaces of possible lemmas and  game structures requiring the application of nested acceleration.
To address this,  our acceleration approach performs symbolic attractor computation with  
``unknown'' lemmas  represented by uninterpreted symbols.
More precisely,  our attractor acceleration technique has two main components.
\begin{itemize}
\item   
In the remainder of this section,  we show how symbolic attractor computation is enhanced with the application of acceleration lemmas as \emph{uninterpreted lemmas}.
An essential part of this computation is the collection of constraints on the unknown lemmas that capture conditions which guarantee  that the lemmas are applicable.
\item
In Section~\ref{sec:resolving}, we describe how we generate instances of the unknown lemmas that meet the accumulated constraints and, therefore, constitute applicable acceleration lemmas.
\end{itemize}

\subsubsection*{Attractor Acceleration with Uninterpreted Lemmas.}

To represent the unknown lemmas in the symbolic attractor computation, 
let us fix a countably infinite set $\lemmasymb$ of triples $(b,st,c)$ of uninterpreted predicate symbols.  The symbols in such a triple represent the corresponding elements of an acceleration lemma. 
In the course of the symbolic computation of attractors, we generate and accumulate constraints, elements of  $\constraints := \FOL{\emptyset}$, featuring these symbols. 
For $\Phi \subseteq \FOLX$, we denote with $\usedlemmas(\Phi)$ the set of triples whose symbols appear in $\Phi$.


\begin{figure}[t]
\begin{algorithm}[H]
\SetAlgoVlined
    \SetKwProg{Fn}{function}{}{}
    \DontPrintSemicolon
    \Fn{\textsc{AttractorAcc}(
		 $\mathcal G = (T,\inputs, \cells, L, \Inv,\delta)$,  
    	 $\mathit{p} \in \{\sys,\env\}$,  
		 $d \in \symstates$)}{
		\setcounter{AlgoLine}{0}
		\nl $a^0$ := $\lambda l.~\bot$;
        $a^1$ := $d$\;
        \nl \For{$n=1,2,\ldots$}{
			\nl \lIf{$a^{n} \equiv_{T} a^{n-1}$}{\Return $a^n$}
			\changed{
			\If{\textsc{Accelerate?}$(G,p,a^n)=\{l\}$}{
                $(\constr, \varphi) := \accreach(\mathcal{G}, p, l, a^n)$\;
				$a^n(l) := a^n(l)\;\lor\;\resolve(\constr, \varphi)$
			}
			}
			\setcounter{AlgoLine}{3}
			\nl $a^{n+1} := a^n \lor \cpre{\mathcal G,p}{a^n}$\; } 		
	}
		 \caption{Accelerated semi-algorithm for computing 
		 $\mathit{Attr}_{\sema{\mathcal G},p}(\sema{d})$.}\label{algo:attractor-acc}	
\end{algorithm}
\vspace{-5mm}
\end{figure}

Algorithm~\ref{algo:attractor-acc} shows the procedure \textsc{AttractorAcc} for attractor computation with acceleration.  
At each iteration, a \emph{heuristic function  \textsc{Accelerate?}} is used to decide whether an acceleration should be attempted at location $l$ at the current step of the computation.  This function returns either $\emptyset$ or a singleton set $\{l\}$ of locations. 
We discuss possible heuristics in Section~\ref{sec:eval}. 

The acceleration consists in computing a formula $\alpha \in \FOL{\cells}$ such that the set $\{(l, \assmt{x}) \mid \assmt{x} \FOLentailsT{T} \alpha\}$ is a subset of the attractor we aim to compute.  
This set is computed in two steps.
\begin{enumerate}[(i)]

\item 
First,  the procedure invokes the function $\accreach: \mathit{GameStructure} \times \{\sys,\env\} \times L \times \symstates \to \constraints \times \FOL{\cells}$.  
$\accreach$ returns a pair of formulas $(\Psi, \varphi)$,  where $\Psi$ is a constraint that captures conditions under which the formula $\varphi$ can be used for acceleration at location $l$.
This is done by constructing the loop game at location $l$ and establishing the conditions for applying an acceleration lemma.  
$\varphi$ represents the conclusion of the acceleration lemma,  which we want to add to $a^n(l)$.
However,  at this point the acceleration lemma is not a concrete one but is represented by uninterpreted predicate symbols which occur in $\varphi$.
Furthermore,  the constraint $\Psi$  over those uninterpreted symbols captures the conditions for lemma application.
The procedure \textsc{AttractorAcc} adds $\varphi$ to the attractor only in case there exists an instantiation of the uninterpreted predicate symbols that satisfies $\Psi$. 
This is done in the next step.
\item Function $\resolve: \constraints \times \FOLX \to \FOL{\cells}$ takes the result from $\accreach$ and searches for an instantiation to the uninterpreted symbols used to represent the unknown acceleration lemmas.  If such exists, it returns the respective instantiation of $\varphi$ that is then added to $a^n(l)$.  Otherwise it simply returns $\bot$, meaning that this acceleration attempt has failed and has no effect.
We defer to Section~\ref{sec:resolving}  the description of the search for instantiations.
\end{enumerate}

Procedure $\accreach$ is described in Figure~\ref{fig:attractor-rec}.  It employs a helper function $\itera: \mathit{GameStructure} \times \{\sys,\env\} \times \constraints \times \symstates \to \constraints \times \symstates$,  which performs symbolic computation of attractor subsets.  $\accreach$ invokes $\itera$ for attractor computation in the loop game (using the $\cpre{}{\cdot}$ operator  for the respective game).
Intuitively, the two procedures apply the method outlined earlier  but additionally collect the constraints for the (nested) lemma applications as those cannot be discharged immediately since the lemmas are uninterpreted.
Note that we cannot use $\textsc{AttractorAcc}$ instead of $\itera$ since the uninterpreted predicates prevent us from identifying if we have reached a fixpoint.
As $\itera$ underapproximates an attractor, it uses the heuristic function \textsc{Iterate?} to determine when to stop the computation and return the current underapproximation.
Furthermore, $\itera$ calls $\accreach$ to perform nested acceleration,  as determined by the heuristic function \textsc{Accelerate?}.

\begin{figure}[t!]
\begin{algorithm}[H]
\SetAlgoVlined
    \DontPrintSemicolon
   \hspace{-.5cm} {\bf function} 
    $\accreach(\mathcal{G}, p, l, d)$\; 
    {\bf pick fresh} $(b, st, c) \in \lemmasymb$\; 
    $\mathcal{G}_{\mathit{loop}} := \loopgame(\mathcal{G}, l, \locend)$ with fresh location $\locend \not\in L$\;
    $d_{\mathit{loop}} := d \cup \{\locend \mapsto st(\cellsnul, \cells)\}$ {\bf where} $\cellsnul = \{e_x \text{: fresh unint. constant}\mid x \in \cells \}$\;
    $(\constr_\mathit{Rec}, a) := \itera(\mathcal{G}_{\mathit{loop}}, p,\top,d_{\mathit{loop}})[\cellsnul \mapsto \cells]$\; 
    $ \constr := (\forall \cells.~\Inv(l) \land b(\cells) \to d(l)) \land (\forall \cells. ~\Inv(l) \land \lnot d(l) \land c(\cells) \to \constr_\mathit{Rec} \land a(l))$\;
\Return  $( \constr, c(\cells) \land \Inv(l))$
\end{algorithm}  
 \begin{algorithm}[H]
\SetAlgoVlined
    \DontPrintSemicolon
   \hspace{-.5cm} {\bf function} 
		 $\itera(\mathcal{G}, p,\constr, d$)\\ 			
		 \lIf{$\neg \textsc{Iterate?}(\mathcal{G}, p,\constr, d)$}{\Return $(\constr, d)$}
		 \If{$\textsc{Accelerate?}(\mathcal{G}, p,\constr, d)=\{l\}$}{
		 $(\constr_\mathit{Rec}, \varphi) := \accreach(\mathcal{G}, p, l, d)$\;
			\Return $\itera(\mathcal G,p, \constr \land \constr_\mathit{Rec}, d[l \mapsto d(l) \lor \varphi])$
		 }
		 \Return $ \itera(\mathcal G,p, \constr,d \lor \cpre{\mathcal{G}, p}{d})$		 
\end{algorithm}
\begin{algorithm}[H]
\SetAlgoVlined
    \DontPrintSemicolon
   \hspace{-.5cm} {\bf function} 
        $\resolve(\constr, \varphi)$\;
		 $\mathit{Symb} := \usedlemmas( \{\constr, \varphi\})$\;
		  {\bf if  find} $\lambda : \mathit{Symb} \to \lemmas(\cells)$ {\bf such that} $\valid{\constr[\lambda]}$
			{\bf then}	\Return $\varphi[\lambda]$ {\bf else} \Return $\bot$ 
\end{algorithm}    
     \vspace{-2mm}
     \caption{Procedures for attractor acceleration using uninterpreted lemmas.}\label{fig:attractor-rec}	
    \vspace{-4mm}
\end{figure}

Next, we describe in detail the function $\accreach$.
Given a location $l \in L$ and symbolic goal $d$, $\accreach$ computes a symbolic state for $l$ from which Player~$p$ can enforce going to $d$, if the uninterpreted lemmas satisfy the computed constraint.
$\accreach$ does so by introducing and symbolically applying an acceleration lemma at location $l$ as follows. 
\begin{itemize}
\item 
    We first pick a fresh triple $(b, st, c)$ of lemma symbols from $\lemmasymb$ for the uninterpreted lemma that we are applying.
\item 
    As in Example~\ref{ex:lemma-loop}, we construct $\mathcal{G}_{\mathit{loop}}$ for $l$, since we want to accelerate at location $l$.
\item 
    We construct $d_{\mathit{loop}}$ obtained from $d$ by mapping the new location $\locend$ to the formula $st(\cellsnul, \cells)$.  
    $st(\cellsnul, \cells)$ expresses the relation between $\cellsnul$, a set of fresh uninterpreted constants with one such constant for each variable in $\cells$, representing the initial assignment, and $\cells$ representing the values of the program variables at $\locend$.
\item 
    We invoke $\itera$ which recursively performs iterations of the attractor computation.
    Intuitively, if Player~$p$ can enforce to reach $d_{\mathit{loop}}(\locend)$, it can either get to the goal $\sema{d}$ or enforce the step relation $st(\cellsnul, \cells)$.     
    As $\itera$ might call $\accreach$, it also returns a constraint $\constr_\mathit{Rec}$.
    The substitution $[\cellsnul \mapsto \cells]$ denotes the syntactic transformation that replaces each $e_x$ by the respective $x \in \cells$.
    This transformation is necessary as, intuitively, for $\constr_\mathit{Rec}$ and $a(l)$ we have to consider for the constraint the initial assignment to $\cells$ in $l$ before performing the step.
\item 
    Finally, $\accreach$ constructs the new constraint $\constr$.
    Intuitively, the first conjunct states that the base of the lemma $b(\cells)$ is included in the goal $d(l)$.  
    The second conjunct states that every state in the conclusion $c(\cells)$ that is not an element of  the goal must be in $\sema{a}$, i.e. either the goal or the step relation is enforceable from $c(\cells)$ if not already at the goal.
    We assume that each quantifier uses a fresh copy (or De Bruijn indexing) of the variables in $\cells$ and applies the appropriate renaming. This ensures the correct replacement of constants in $E$ with variables as described in the previous step.
\item 
    We return the new constraint $\constr$ and the conclusion $c(\cells)$ of the applied lemma.
    The constraint $\constr$ ensures, together with the property of an acceleration lemma, that from $c(\cells)$, Player~$p$ can enforce reaching $d$.
\item
    The incompleteness of $\itera$ is not a problem, as terminating early with an underapproximation will only result in a stronger constraint.
\end{itemize}

What remains is to define the function $\resolve(\constr, \varphi)$ that searches for an instantiation of the  uninterpreted symbols, representing the used acceleration lemmas,  that satisfies $\constr$ and defines valid lemmas.  
If such an instantiation is found,  it returns a formula $\varphi$ in which the terms applying these symbols have been replaced with formulas in \FOL{\cells}.  In Section~\ref{sec:resolving} we discuss the practicalities of the check for the existence of lemmas.  Function $\resolve$,  shown in Figure~\ref{fig:attractor-rec},  requires finding a mapping  $\lambda : \usedlemmas(\{\constr, \varphi\}) \to \lemmas(\cells)$ of the lemma symbol triples used in $\constr$ and $\varphi$ to the set of lemmas $\lemmas(\cells)$. 
For $\psi \in \{ \constr, \varphi\}$ we denote with $\psi[\lambda]$ the formula obtained from $\psi$ by the following transformation. 
For each  $(b,st,c)$  such that $\lambda((b,st,c)) =(\base,\step,\conc)$,  replace each predicate term of the form
\begin{itemize}
\item  $b(\{t_x\}_{x\in\cells})$  by $\base[x \mapsto t_x \mid x \in \cells]$;
 $c(\{t_x\}_{x\in\cells})$ by $\conc[x \mapsto t_x \mid x \in \cells]$;
\item $st(\{t_x\}_{x\in\cells},\{t'_x\}_{x\in\cells})$ by $\step[x \mapsto t_x, x' \mapsto t'_x \mid x \in \cells]$.
\end{itemize}
By the construction of $\constr$ and $\varphi$ in $\accreach$,  all the terms containing the lemma predicate symbols as top symbols are of the above form,  and hence, no uninterpreted acceleration lemma symbols appear in $\psi[\lambda]$. 
Furthermore,  all uninterpreted constants introduced during the construction have been mapped back to variables and do not appear in the result returned to \textsc{AttractorAcc}.

Note that since $\accreach$ only performs symbolic computation and collects constraints,  it always returns a result.
If there exists no instantiation of the uninterpreted predicate symbols that satisfies the constraints, function $\resolve$ will fail to find one,  in which case it will return $\bot$. 

\begin{example}\label{ex:show-accel}
Consider our running example from Figure~\ref{fig:ex-motivating} where Player~\sys\ has to enforce reaching $l_G$.
Suppose we accelerate in Algorithm~\ref{algo:attractor-acc} at location $l_0$ after the first iteration, invoking procedure $\accreach(\mathcal{G}, \sys, l_0, d)$ with $d = a^2 = \{ l_0 \mapsto (x \leq 42), l_G \mapsto \top \}$.
$\accreach$ picks $(b, st, c) \in \lemmasymb$ and constructs $\mathcal{G}_{\mathit{loop}}$ as shown in Figure~\ref{fig:ex-loop-game}.
It then calls $\itera(\mathcal{G}_{\mathit{loop}}, \sys,\top,d_{\mathit{loop}})$ with $d_{\mathit{loop}}:=\{ l_0 \mapsto (x \leq 42), l_G \mapsto \top, \locend \mapsto st(e_x,x)\}$, 
where $e_x$ is a fresh constant for $x \in\cells$.  
Suppose $\itera$ performs one iteration before $\textsc{Iterate?}$ tells it to terminate with $(\top, \{l_0 \mapsto \varphi, l_G \mapsto \top,\locend \mapsto st(e_x,x)\})$ where
$\varphi\equiv_{T}  \forall i.~x \leq 42 \lor  i=0 \lor st(e_x,x-i) \lor st(e_x,x+i)$. 
Back in procedure $\accreach$,  we have $a(l_0) \equiv_{T} \varphi[e_x \mapsto x] \equiv_{T}
\forall i.~x \leq 42 \lor  i=0 \lor st(x,x-i) \lor st(x,x+i)$
and $\constr := (\forall x.~ b(x) \to x \leq 42) \land ( \forall x.~ x > 42 \wedge c(x) \to  a(l_0))
\}$.   
Hence, $\accreach$ returns $(\constr,c(\cells))$.
Then in $\resolve$, it is easy to see that for $\lambda((b,st,c)) := (x \leq 42, x' < x, \top)$ the constraint $\constr$ is valid. 
Applying $\lambda$ to $c(\cells)$ yields $\top$ as result.
Back in $\textsc{AttractorAcc}$ we get $a^1(l_0) := a^1(l_0) \lor \top$ and terminate.
\end{example}

\begin{theorem}[Soundness of Algorithm~\ref{algo:attractor-acc}]\label{thm:attractor-acc}
Let $\mathcal{G}$ be a reactive program game structure,   $p \in \{\sys,\env\}$,  and $d \in \symstates$.
If $\textsc{AttractorAcc}(\mathcal G, p,d)$ terminates returning $a$, then it holds that $\sema{a} = \mathit{Attr}_{\sema{\mathcal G},p}(\sema{d})$. 
\end{theorem}

 Theorem~\ref{thm:attractor-acc}~states the correctness of our acceleration procedure.
It follows from the property of functions $\accreach$  and $\resolve$ stated in the lemma below which, intuitively, states that the constraints only allow applicable lemmas to be instantiated.
We prove Theorem~\ref{thm:attractor-acc} and Lemma~\ref{lem:correcness-accA} in Appendix~\ref{app:proofs}.

\begin{lemma}\label{lem:correcness-accA}
Let $\mathcal{G}$ be a reactive program game structure,  $p \in \{\sys,\env\}$,  $l \in L$ and $d \in \symstates$. 
Then, if $\accreach(\mathcal G, p,l,d)$ returns the pair $(\constr,\varphi)$, then for every mapping $\lambda: \usedlemmas(\{\constr, \varphi\}) \to \lemmas(\cells)$ such that the formula $\constr[\lambda]$ is valid it holds that $\sema{\{ l \mapsto \varphi[\lambda]\}} \subseteq \mathit{Attr}_{\sema{\mathcal G},p}(\sema{d})$.
\end{lemma}

{\it Note:} 
Our method is inherently incomplete and it is difficult to characterize the class of infinite-state games on which it terminates.
This is not surprising, as pointed out in~\citep{AbdullaBd08} transferring decidability results from verification to games is in general non-trivial.

\subsection{Extracting Reactive Programs from Acceleration Lemmas}\label{sec:strategy-extraction-accel}
The extraction of programs in the presence of acceleration follows the same principles as in Section~\ref{sec:strategy-extraction}. We only need to define the sub-program constructed for the application of acceleration during the attractor computation, that is the invocation of procedure \accreach.

When extracting a program for \accreach, we add new labels in the GOTO program for each location in the loop game $\mathcal{G}_{\mathit{loop}}$.
Note that those new labels are different from the labels corresponding to the locations in the original game $\mathcal{G}$.
The program labels $l_0^L$,  $l_G^L$ and $l_E^L$ in Figure~\ref{fig:ex-program} result from the loop game in Figure~\ref{fig:ex-loop-game},  constructed by \accreach.
Before starting the extraction for \itera, we introduce an auxiliary copy of the program variables that stores the  values of $\cells$ at the label $l$ (this corresponds to the set of fresh variables $E$).
Storing these values is necessary in order to check conditions on the step relation in the program for \itera.
The program in Figure~\ref{fig:ex-program} contains such an auxiliary variable  $x_0$ and the copy assignment $x_0 := x$ at label $l_0^L$.

For \itera\ we extract the program as outlined in Section~\ref{sec:strategy-extraction} except that the conditions in the program might contain uninterpreted symbols for the lemmas.
If in the program \itera\ parts of the original attractor are reached, we add a jump back to the respective location of $\mathcal{G}$.
If $\locend$ is reached, the program jumps back to the label of $l$. 
If \itera\ applies acceleration (invokes \accreach) we apply this extraction process recursively. 

The result is a program skeleton from which we still have to remove the uninterpreted lemma symbols.
After \resolve\ has found and instantiation with concrete lemmas,  we use this instantiation to replace all occurrences of uninterpreted lemma symbols in the conditions of the extracted program.
In our running example, when generating the program in Figure~\ref{fig:ex-program},  we will first get a program including statements of the form\ \texttt{if($\step(x_0, x + i)$)\{ $x$:=$x+ i$; \textcolor{purple}{goto} $\color{brown}{l_{E}^L}$\}}. 
Then, we plug in the acceleration lemma from Example~\ref{ex:lemma} and $\step(x_0, x + i)$ becomes $x + i < x_0$ (as $x + i$ corresponds to $x'$ in $\step$ and $x_0$ to $x$) which is exactly the statement that we have in Figure~\ref{fig:ex-program}.

\subsection{Acceleration Beyond Attractors}\label{sec:accel-buechi}

The symbolic semi-algorithm for solving Büchi games (and, by duality,  co-B\"uchi games) outlined in Section~\ref{sec:solving} is accelerated by using $\textsc{AttractorAcc}$ instead of $\textsc{Attractor}$. 
This amounts to replacing the calls in lines 4 and 5 in Algorithm~\ref{algo:buchi} by calls to $\textsc{AttractorAcc}$.
This accelerates the computation of attractor sets in the solving procedure, and helps the overall computation of the set of winning states to converge. 
More concretely,  at line 4 $\textsc{AttractorAcc}$ accelerates the computation of the set of states from which the B\"uchi player can reach an accepting location.
At line 5,  the acceleration is applied for the co-B\"uchi player,  accelerating the computation of the set of states winning for the  co-B\"uchi player.
For the B\"uchi game in Example~\ref{ex:buchi-motivating} the procedure for solving B\"uchi games with accelerated attractor computation successfully terminates.
The evaluation in Section~\ref{sec:eval} further demonstrates the usefulness of attractor acceleration for solving B\"uchi games.

We now show an example where the convergence of $\textsc{AttractorAcc}$ is insufficient to ensure the convergence of the procedure for Büchi games.
Consider the reactive program game structure in Figure~\ref{fig:game-cobuchi} and the Büchi objective for Player \sys\ defined by $B:=\{l_0\}$.
That is,  the objective of Player~\sys\ is to visit location $l_0$ infinitely often.
The table in Figure~\ref{fig:table-cobuchi} shows the computation of the procedure in Algorithm~\ref{algo:buchi}.
We observe that the sets $w_{1-p}^i$ will keep growing, and the computation never terminates.
Player \env\ wins the game for every possible state since the initial value of $x$ bounds the number of visits to $l_0$.
However, this argument cannot be captured with attractor acceleration, as Player \env\ \emph{cannot force} Player \sys\ to decrease $x$. 

\begin{figure}[t]
    \begin{subfigure}[b]{0.45\textwidth}
         \centering
			\scalebox{0.9}{\begin{tikzpicture}[->,>=stealth',shorten >=1pt,auto,node distance=2.5cm]\footnotesize
  \tikzstyle{every state}=[fill=none,draw=black,text=black,inner sep=1.5pt, minimum size=12pt,thick,scale=0.8]

  \node[state, accepting] (s0)     {$l_0$};
  \node[state] (s1) [right of=s0, xshift=1.5cm] {$l_1$};
  \node[state] (s2) [below of=s0,yshift=.2cm] {$l_2$};
   \node[state] (s3) [below of=s1,yshift=.2cm] {$l_3$};

   \node[fill=black,draw=black,minimum size=0.5pt] (e01) at (1,0) {};
   \node[fill=black,draw=black,minimum size=0.5pt] (e12) at (1.8,-.8) {};
   \node[fill=black,draw=black,minimum size=0.5pt] (e20) at (0,-.9) {};
   \node[fill=black,draw=black,minimum size=0.5pt] (e23) at (1.8,-1.85) {};
   \node[fill=black,draw=black,minimum size=0.5pt] (e11) at (4,0) {};
   \node[fill=black,draw=black,minimum size=0.5pt] (e33) at (4,-1.85) {};
  
  \path (s0) edge  node[above] {$\top$} (e01);
  \path (e01) edge  node[above] {$\mathit{skip}$} (s1);
  \path (s1) edge[bend left=10]  node[above] {$\top$} (e11);
   \path (e11) edge[bend left=10]  node[below] {$\mathit{skip}$} (s1);
  \path (s1) edge  node[sloped] {$\top$} (e12);
  \path (e12) edge  node[sloped] {$x:=x-1$} (s2);  
  \path (s2) edge  node {$x\leq 0$} (e23);
  \path (e23) edge  node {$\mathit{skip}$} (s3);  
  \path (s2) edge  node[sloped] {$x>0$} (e20);
  \path (e20) edge  node[sloped] {$\mathit{skip}$} (s0);  
  \path (s3) edge[bend left=10]  node[above] {$\top$} (e33);
   \path (e33) edge[bend left=10]  node[below] {$\mathit{skip}$} (s3);

\end{tikzpicture}}
			\vspace{-.2cm}			
         \caption{Game with integer program variable $x$.}
         \label{fig:game-cobuchi}
     \end{subfigure}
     \hfill
     \begin{subfigure}[b]{0.54\textwidth}
         \centering
\scalebox{0.9}{%
		\begin{tabular}{l|l|l|l|l||l|l|l} 
 & $f^0$ &  $a^0$     & $w_{1-p}^1$ &
     $f^1$ &  $a^1$    & $w_{1-p}^2$  & ...\\
 \hline\hline
$l_0$ & 
$\top$ &  $\top$     & $\textcolor{blue}{x\leq 1}$ &
$x>1$ &  $x>1$     & $\textcolor{blue}{x \leq 2}$  & ...\\
\hline
$l_1$ & 
$\bot$ &  $x>1$    & $x \leq 1$ &
$\bot$ &  $x>2$    & $x \leq 2$  & ...\\
\hline
$l_2$ & 
$\bot$ &  $x>0$    & $x\leq 1$ &
$\bot$ &  $x>1$    & $x \leq 2$  & ...\\
\hline
$l_3$ & 
$\bot$   &  $\bot$   & $\top$ &
$\bot$   &  $\bot$   & $\top$  & ...\\
\hline
\end{tabular}}
         \caption{Sets computed during the procedure.}
         \label{fig:table-cobuchi}
     \end{subfigure}
     
     	\vspace{-3mm}
\caption{Example demonstrating the need for acceleration of the computation of  the set  $w_{1-p}^n$ in the procedure for solving B\"uchi games outlined in Section~\ref{sec:solving}.}\label{fig:acc-cobuchi}
\vspace{-5mm}
\end{figure}

More abstractly,  attractor acceleration cannot sufficiently enlarge the set $w_{1-p}^i$ to enforce convergence of the outer loop of the procedure for B\"uchi games. 
The reason is that attractor acceleration cannot account for states in the winning region of Player~$1-p$ from which
\begin{itemize}
\item Player~$1-p$ cannot enforce reaching $w_{1-p}^i$,  and
\item the repeated visit of a B\"uchi accepting location by Player~$p$ will eventually lead to $w_{1-p}^i$.
\end{itemize} 

Motivated by this observation, we introduce an additional acceleration method, called \emph{co-Büchi acceleration}, that extends the scope of acceleration-based symbolic game solving.
Intuitively, this method allows us to establish arguments of the form:
``\textbf{If} Player~$p$ with the Büchi objective keeps visiting \emph{a given B\"uchi  accepting location}, \textbf{then} the set of winning states of Player~$1-p$ must be reached eventually''.
The aspect of ``reaching the winning states eventually'' is captured by the usage of acceleration lemmas.
However, in contrast to attractor acceleration, we will not use that the step relation is enforceable by a player, but rather that it is unavoidable by Player~$p$ upon revisiting the  given accepting location. 

Before diving into the detailed description,  we must highlight that similarly to attractor acceleration, the co-Büchi acceleration is focused on a single accepting location.
However, as the set $B$ of Büchi accepting locations might contain more than one  location, we need to make sure that we do not incorrectly exclude states from the winning region of Player~$p$ by accounting only for repeated visits of the given location. 
Hence,  in contrast to attractor acceleration,  which performs attractor computation in a loop game,  for co-Büchi acceleration we consider a loop game with a B\"uchi winning condition  for which we compute an underapproximation of the winning region of the co-B\"uchi player.
Thus,  co-Büchi acceleration is not the dual of attractor acceleration,  but an extension.\looseness=-1

The overall procedure is formalized in Figure~\ref{fig:buchi-acc}. 
The game-solving procedure \textsc{SolveB\"uchi} uses $\acccobuchi$ to employ co-Büchi acceleration to accelerate the outer fixpoint, which computes the sets $f^1,f^2,\ldots$, in addition to invoking \textsc{AttractorAcc}.
On a high level, the co-Büchi acceleration computation focuses on one of the B\"uchi accepting locations $l \in B$ and tries to establish a set of states with location $l$ from which the Büchi Player~$p$ cannot revisit $l$ infinitely often, and cannot win the game by visiting some of the locations in $B \setminus \{l\}$ infinitely often. Such states are winning for Player~$1-p$ and can be added to $w_{1-p}^n$.
To establish that Player~$p$ cannot visit $l$ infinitely often, we use an acceleration lemma and a loop game but require Player~$p$ to enforce the negation of the step relation upon revisiting $l$.   From states where this is impossible,  Player~$p$ cannot enforce visiting $l$ infinitely often.  As explained earlier, to account for the remaining B\"uchi accepting locations $B \setminus \{l\}$,  the loop game constructed by $\acccobuchi$ has a B\"uchi winning condition for Player~$p$.

The function $\acccobuchi : \mathit{GameStructure} \times \{\sys,\env\} \times L \times \symstates  \to \constraints \times \FOL{\cells}$ generates the constraints for the acceleration lemmas.
Therefore, it constructs a loop game for a location $l \in B$, the set of B\"uchi accepting locations, and equips it with a B\"uchi winning condition $f_{\mathit{loop}} := f \cup \{\locend \mapsto \neg st(\cellsnul, \cells)\}$. Intuitively,  $f_{\mathit{loop}}$ requires that in the loop game, the B\"uchi player enforces visiting infinitely often a location in  $B\setminus\{l\}$, or a visit to $\locend$ while also ensuring that the step relation of the applied lemma is violated.  
Thus,  the winning region of the co-B\"uchi player in this game are states from which it can enforce that the play does not visit $B \setminus \{l\}$ infinitely often and \emph{if} it visits location $\locend$, \emph{then} the step relation must be satisfied at location $\locend$. 
If Player~$p$ loses this loop game, Player~$p$ can only win the overall game by visiting the location $l$ infinitely often while applying $\step$, which is not possible. This allows extending the winning region of Player~$1-p$ using an acceleration lemma. Similarly to the function $\accreach$, if $ \resolve$ succeeds, the procedure adds the set of states represented by the conclusion of the lemma to the set $w_{1-p}^n(l)$.

The function $\iterb:\mathit{GameStructure} \times \{\sys,\env\} \times \constraints \times \symstates \times \symstates \to \constraints \times \symstates$ performs the steps of the procedure for B\"uchi games and invokes $\itera$ for the (accelerated) attractor computation. A key detail here is that the attractor sets computed for the B\"uchi player must not be underapproximated.  
This is required as the complement set is added to the winning region of the opponent, which would be incorrect in case of an  underapproximation.
To this end, we add an additional constraint defined as $\precise(\mathcal{G}, p,a) := \forall l \in L.~\cpre{\mathcal{G}, p}{a}(l) \to a(l)$,  which states that the corresponding symbolic set overapproximates the attractor.

Similarly to Theorem~\ref{thm:attractor-acc} we obtain the following correctness property.
\begin{proposition}[Soundness of \textsc{SolveB\"uchiAcc}]\label{thm:buechi-acc}
Let $\mathcal{G}$ be a reactive program game structure, $p \in \{\sys,\env\}$, and $B \subseteq L$.
If $\textsc{SolveB\"uchiAcc}(\mathcal G, p, \{l \mapsto \Inv(l) ~|~ l \in B \})$ terminates returning $w$, then it holds that $\sema{w} = W_{1-p}(\sema{\mathcal G},\cobuchi(B))$.
\end{proposition}

\begin{figure}[t]
\begin{algorithm}[H]
\SetAlgoVlined
    \SetKwProg{Fn}{function}{}{}
    \DontPrintSemicolon
    \Fn{\textsc{SolveB\"uchiAcc}(
		 $\mathcal G = (T,\inputs, \cells, L, \Inv,\delta)$,  
    	 $\mathit{p} \in \{\sys,\env\}$,  
		 $f \in \symstates$)}{
		\setcounter{AlgoLine}{0}
		\ldots\;
\changed{		 \setcounter{AlgoLine}{3}
	     \nl $a^n := \textsc{AttractorAcc}(\mathcal G, p, f^n)$\;
	     \nl $w^{n}_{1- p} := \textsc{AttractorAcc}(\mathcal G,1- p,\neg  \cpre{\mathcal G,p}{a^n})$\;
	     \If{\textsc{AccelBuchi?}$(G,p,w^n_{1-p})=\{l\}$}{
	     $(\Psi,\varphi) := \acccobuchi(\mathcal{G}, p, l, f^n \wedge (\Inv \land \neg w^{n}_{1- p}) )$\;
		 $w_{1-p}^n(l) := w_{1-p}^n(l)  \lor \resolve(\Psi,\varphi)$
		 }
		 }
		\setcounter{AlgoLine}{5}
		 \nl $f^{n+1} :=  f^{n+1} \wedge (\Inv \land \neg w^{n}_{1- p})$\;     
	}
\end{algorithm}\label{alg:acc-buechi}
\begin{subfigure}[b]{0.43\textwidth}
\begin{algorithm}[H]
\SetAlgoVlined
    \DontPrintSemicolon
   \hspace{-.5cm} {\bf function} 
    $\acccobuchi(\mathcal{G}, p, l, f)$\;
     {\bf pick fresh} $(b, st, c) \in \lemmasymb$\;
  $\mathcal{G}_{\mathit{loop}} := \loopgame(\mathcal{G}, l, \locend)$ with fresh location $\locend$\;
  $f_{\mathit{loop}} := f \cup \{\locend \mapsto \neg st(\cellsnul, \cells)\}$ \textbf{where} $\cellsnul = \{e_x \text{: fresh unint. constant}\mid x \in \cells \}$\;
$(\constr_\mathit{Rec}, a) := \iterb(\mathcal{G}_{\mathit{loop}}, p,\emptyset,f_{\mathit{loop}})[\cellsnul \mapsto \cells]$\;
$ \constr := (\forall \cells.~\Inv(l) \land b(\cells) \to \lnot f(l))$, 
$ \land (\forall \cells. ~\Inv(l) \land f(l) \land c(\cells)$  \hspace{5mm} $\to \constr_\mathit{Rec} \land a(l))$\;
\Return  $( \constr, c(\cells) \land \Inv(l))$
\end{algorithm}  
\end{subfigure}
\hfill
\begin{subfigure}[b]{0.56\textwidth}
\begin{algorithm}[H]
\SetAlgoVlined
    \DontPrintSemicolon
   \hspace{-.5cm} {\bf function} \textsf{IterB}(
		 $\mathcal G,\mathit{p},\constr,f,w_{1-p}$)\\ 			
		 \lIf{$\neg \textsc{Iterate?}(\mathcal{G}, p,\constr, f, w_{1-p})$}{\Return $(\constr, w_{1-p})$}
		 $(\constr_1,a) := \itera(\mathcal G, p, \constr,f)$\;
		 $ \constr_1 := \constr_1 \land \precise(\mathcal{G}, p,a)$\; 
	     $(\constr_2, w'_{1- p}) := \itera(\mathcal G,1- p,\Inv \land \neg  \cpre{\mathcal G,p}{a})$\;
	     $f' :=  f \wedge (\Inv \land \neg w'_{1- p})$\;  
		 \If{$\textsc{AccelBuchi?}(\mathcal{G}, p,\constr, f', w_{1-p}')=\{l\}$}{
		 $(\constr_3, \varphi) := \acccobuchi(\mathcal{G}, p, l, f')$\;
		 $f' := f'[l\mapsto f'(l) \land \neg \varphi]$\; 
         $w'_{1-p}:= w'_{1-p}[l \mapsto w'_{1-p}(l) \lor \varphi]$\;
			\Return $\iterb(\mathcal G,p,\constr\land\constr_1\land\constr_2\land\constr_3, f',w'_{1-p})$
		 }
 \Return $ \iterb(\mathcal G,p, \constr \land \constr_1 \land \constr_2,f', w'_{1-p})$		 	
\end{algorithm}
\end{subfigure}
     \vspace{-4.5mm}
\caption{Accelerated semi-algorithm  for solving  B\"uchi games.}
\label{fig:buchi-acc}
     \vspace{-4.5mm}
\end{figure}


\section{Synthesis of Acceleration Lemmas and Programs}\label{sec:resolving}
In this section we describe how our acceleration method instantiates the uninterpreted predicate symbols representing acceleration lemmas by actual acceleration lemmas.
In order to ensure that the instantiations constitute valid acceleration lemmas, and also in order to constrain the search space,  the lemmas are derived from templates.
The templates associate with the individual lemma components parameterized  formulas conforming to certain syntactic restrictions.
By replacing the uninterpreted predicates representing the lemmas by the respective templates,  we restrict the shape of the sought lemmas.  
The lemma instantiation task is then to find values for the template parameters, called meta-variables,  that satisfy the constraints generated during the symbolic attractor computation. 
However,  single instantiations of the meta-variables may be overly specific and thus result in acceleration steps with negligible effect.
We address this issue by proposing a method based on quantifier elimination that computes the combined acceleration effect of all instantitations to the meta-variables that satisfy the constraints. 
We present this method in Section~\ref{sec:qe} below,  but first we provide in  Section~\ref{sec:templates} the definition of the templates to which our lemmas must conform.


\subsection{Acceleration Lemma Templates}\label{sec:templates}
The derived lemmas must fulfill two requirements: they must satisfy the constraint generated during symbolic acceleration, and they must fulfill the conditions of  Definition~\ref{def:lemma}. 
Since the constraint is a \FOLX\ formula in some theory $T$, our goal is to formulate the lemma generation task as an SMT problem.  
We use templates representing acceleration lemmas in a fragment of linear mixed integer-real arithmetic.  
In principle,  the class of templates can be extended to other theories. 
Furthermore,  our method also allows for using custom lemmas.
We need to define the templates  used for lemma generation in a way that ensures that the results are valid w.r.t. the second-order condition in Definition~\ref{def:lemma}. 
In order to do this, we make use of the following proposition.
\begin{proposition}\label{prop:ranklemma}
    Let $\epsilon > 0$ be a constant, and let $r \in \funcTerms$ be a function term of real or integer type over variables $V \subseteq \vars$.
    Then, $(r(V) \leq 0, r(V') + \epsilon \leq r(V), \top)$ is an acceleration lemma over $V$.
    
    We call $r$ the \emph{ranking function}.
\end{proposition}
The above property provides us with a way to generate acceleration lemmas that satisfy the first condition in Definition~\ref{def:lemma}.
To ensure that we can effectively check satisfiability of the generated constraints, a meaningful class of ranking functions are affine functions with bounded coefficients. 
For instance,  when $\cells_\mathit{Num}$ are the program variables with a numerical type,  we consider 
\[r = \sum_{x \in \cells_\mathit{Num}} p_x x + d \text{ with } d \in \Real \text{ and } p_x \in \{-1,0,1\}.\] 

Such a lemma is typically applicable only in some part of the state space. 
This necessitates the application of multiple lemmas and specifying some properties to be invariant under the step relation. 
Hence, our templates should allow adding invariants to lemmas as stated in the following.
\begin{proposition}\label{prop:invlemma}
    Let $(\base, \step, \conc)$ be an acceleration lemma over $V$ and let $\inv \in \FOL{V}$ be a formula.  Then,
    $(\base \land \inv, \step \land \inv[V \mapsto V'], \conc \land \inv)$ is also an acceleration lemma.
    
    We call $\inv$ a \emph{lemma invariant}. \footnote{\emph{Lemma invariants} are generated by our acceleration procedure and are not related to the \emph{location invariants} that are part of reactive program game model.}
\end{proposition}
We define templates for lemma  invariants  as linear inequalities with bounded coefficients.
More concretely,  we consider linear inequalities with $B$ variables for some selected bound $B \in \Nat$, i.e.,
\[\inv_B = \sum_{x \in \cells_\mathit{Num}} p_x x \leq d \text{ with } d \in \Real, p_x \in \{-1,0,1\} \text{ and } \sum_{x \in \cells}|p_x| \leq B.\]
 We distinguish the special case where $B = 1$, which results in the simpler set of lemma invariant templates $\inv_T : = \{x \leq d, x \geq d, x = d \mid x \in \cells_\mathit{Num}, d \in \Real \}$. 
As lemma invariants need not satisfy any additional restrictions, we can allow conjunctions of inequalities, or even other given predicates. 
\smallskip

Putting everything together,  the lemma template that we  use is the one from Proposition~\ref{prop:ranklemma} combined according to Proposition~\ref{prop:invlemma} with a lemma invariant $\inv$ that is a conjunction of several instances of $\inv_B$ and $\inv_T$ where each part of the template has its unique set of meta-variables. 

\subsection{Lemma Instantiation via Quantifier Elimination}\label{sec:qe}

It remains to discuss how to transform the templates into a mapping  
$\lambda: \mathit{Symb} \to \lemmas(\cells)$ of the used lemma symbols  $\mathit{Symb} := \usedlemmas( \{\constr, \varphi\})$ to actual lemmas.
Recall that $\lambda$ is required by function \resolve\ in Figure~\ref{fig:attractor-rec} in order to compute the acceleration result.

We define a \emph{template mapping} to be a function $\tau: \mathit{Symb} \to \FOL{\cells \cup \mathit{Meta}}\times\FOL{\cells \cup \mathit{Meta}}\times \FOL{\cells \cup \mathit{Meta}}$ that maps the uninterpreted lemma symbols to lemma templates containing meta-variables $\mathit{Meta}$ (like $p_x$ or $d$ from above).  
For an instantiation of the meta-variables $M: \mathit{Meta} \to \funcTerms$, that maps the meta-variables to terms without meta-variables, we define $\lambda_{\tau, M}: \mathit{Symb} \to \lemmas(\cells)$ as $\lambda_{\tau, M}(s) = \tau(s)[M]$ where in the image of $\tau$ all meta-variables $m \in \mathit{Meta}$ are simultaneously replaced by their respective terms $M(m)$ (slightly abusing the notation $\cdot[\cdot]$).

Thus, given a constraint $\constr$ and a formula $\varphi \in \FOL{\cells}$ with uninterpreted lemma symbols, the task of finding a mapping $\lambda$ reduces to finding a model $M: \mathit{Meta} \to \funcTerms$ of the formula $\constr[\tau(\mathit{Meta})]$,  where 
$\constr[\tau(\mathit{Meta})]$ is obtained from $\constr$ by applying $\tau$ as substitution.
If a model $M$ is found (by the SMT solver)  we have that $\constr[\lambda_{\tau, M}]$ evaluates to true, and the function $\resolve$ then returns $\varphi[\lambda_{\tau, M}]$.
For program extraction,  $M$ provides a concrete instantiation of the meta-variables, and hence,  a concrete acceleration lemma that we use as described in Section~\ref{sec:strategy-extraction-accel}.

However, the generated model might yield an acceleration lemma that is not general enough.  
For example, recall the acceleration lemma used for  our running example in Example~\ref{ex:show-accel}, but suppose 
that now we want to use the method outlined above to generate an acceleration lemma using our template format.
As our template allows for  the use of lemma invariants,   a model generated by the SMT solver might result in an acceleration lemma of the form $(x \leq 42, x' < x \land x' \leq 100,  x \leq 100)$ which is the original lemma with the additional lemma invariant $x \leq 100$.  Unlike the lemma in Example~\ref{ex:show-accel}, this lemma does not result in immediately reaching a fixpoint in the attractor computation. 
One possible way to mitigate this, is to enumerate models and thus generate multiple acceleration lemmas that can be all applied in \accreach\ to further extend the computed attractor. 

%

Instead of computing instantiations of the meta-variables one by one from the satisfying assignments of $\constr[\tau(\mathit{Meta})]$, we can take a different approach. 
We can consider a formula in which the meta-variables are explicitly existentially quantified:
\[ \Phi(\cells) := \exists \mathit{Meta}.~\constr[\tau(\mathit{Meta})] \land \varphi[\tau(\mathit{Meta})].\]
The set of assignments that satisfy $\Phi(\cells)$ consists of those assignments to $\cells$ for which there exists an  instantiaton of the meta-variables (that is, a valid acceleration lemma).  
Thus,  it implicitly captures all the acceleration lemmas that are possible results of \resolve.
Applying \emph{quantifier elimination} to $\Phi(\cells)$,  we obtain a formula $\mathsf{QElim}(\Phi)$ that characterizes the final conclusion,   which can be seen as the union of the conclusions of all acceleration lemmas satisfying the templates and the constraints.  In our running example, we obtain $\mathsf{QElim}(\Phi) \equiv \top$, as there exist an instantiation of the meta-variables that results in a lemma with conclusion $\top$ (for example the one from our example).
With this approach,  \resolve\ returns $\mathsf{QElim}(\Phi)$ as the result of the acceleration.


\subsection{Program Extraction via Skolem Function Synthesis}

The formula $\mathsf{QElim}(\Phi)$ characterizes all possible results of \resolve\ and is sufficient for computing a symbolically represented set of states to add to the  attractor.
However,   for extracting a program as in Section~\ref{sec:strategy-extraction-accel}, we need concrete acceleration lemmas, i.e.  concrete values for $\mathit{Meta}$.  
Since for different assignments satisfying $\mathsf{QElim}(\Phi)$, the concrete values of $\mathit{Meta}$ can be different,  our goal is to generate
functions for the meta-variables with arguments in $\assignments{\cells}$.
For each $\assmt{x}$ satisfying $\mathsf{QElim}(\Phi)$ these functions should yield a valid lemma.
We consider the  formula
\[\forall \cells.\exists \mathit{Meta}.~
\mathsf{QElim}(\Phi) \to  (\constr[\tau(\mathit{Meta})]\land \varphi[\tau(\mathit{Meta})]).\]
and the task becomes to compute \emph{Skolem functions} $\mathit{SK}_m : \assignments{\cells} \to \mathit{Domain}(m)$ for the meta-variables $m \in \mathit{Meta}$.
This is essentially a functional synthesis problem.
One way to do that is to skolemize the above formula by substituting $\mathit{Meta}$ using uninterpreted function symbols and then solve the resulting second-order satisfiability problem.

A tuple of Skolem functions  $\mathit{SK}_m$, one for each $m \in \mathit{Meta}$, characterizes a lemma as a function of  the values of $\cells$ at the point \emph{when we start applying acceleration}.
Our concrete lemmas are therefore given by the instantiation $\lambda_{\tau, m \mapsto \mathit{SK_m}(\cells_{SK})}$ where $\cells_{SK}$ is a copy of $\cells$ storing the values of the program variables before entering the sub-program generated for acceleration.
Aside from these additional auxiliary variables, the program generation proceeds as described in Section~\ref{sec:strategy-extraction-accel}.



\section{Evaluation}\label{sec:eval}
\begin{table}[t!]
\centering
\scalebox{0.68}{
\begin{tabular}[t]{l|cccccc||cr|rrrr||r}
    Name & $\Omega$ & ST & $|L|$ & $|\cells|$ & $|\inputs|$ & Win & A & rp~ & R~ & T~ & G~ & M~ & extr \\
\hline
    Cinderella 1.4 \cite{cinderella,BodlaenderHKSWZ12}
                   & S & \Real & 4 & 5 & 5 & \env & 0 & 0.3 & MO & NI & \best{0.2} & TO &   - \\
    Cinderella 1.5 & S & \Real & 4 & 5 & 5 & \env & 0 & 0.4 & MO & NI & \best{0.2} & TO &   - \\
    Cinderella 1.6 & S & \Real & 4 & 5 & 5 & \env & 0 & 0.4 & MO & NI & \best{0.2} & TO &   - \\
    Cinderella 1.8 & S & \Real & 4 & 5 & 5 & \env & 0 & 0.5 & MO & NI & \best{0.3} & TO &   - \\
    Cinderella 1.9 & S & \Real & 4 & 5 & 5 & \env & 0 & 0.9 & MO & NI & \best{0.6} & TO &   - \\
    Cinderella 2.0 & S & \Real & 4 & 5 & 5 & \sys & 0 & \best{0.5} & MO & NI & 0.6 & TO & 0.7 \\ 
    Cinderella 2.5 & S & \Real & 4 & 5 & 5 & \sys & 0 & \best{0.5} & MO & NI & \best{0.5} & TO & 0.7 \\
    Cinderella 3.0 & S & \Real & 4 & 5 & 5 & \sys & 0 & 0.5 & MO & NI & \best{0.4} & TO & 0.7 \\
\hline
    Box~\cite{NeiderT16} 
                & S & \Int & 4 & 2 & 4 & \sys & 0 & \best{0.2} &     &     & \best{0.2} & 3.9 & 0.4 \\
    Box Limited & S & \Int & 4 & 2 & 2 & \env & 0 & 0.3 & 0.9 &  TO & \best{0.1} & 2.3 &   - \\
    Diagonal    & S & \Int & 4 & 2 & 2 & \sys & 0 & 0.2 & 4.9 &  MO & \best{0.1} & 3.4 & 0.3 \\
    Evasion     & S & \Int & 4 & 4 & 4 & \sys & 0 & \best{0.2} & 2.3 &  TO & 0.3 &  57 & 0.5 \\
    Follow      & S & \Int & 4 & 4 & 4 & \sys & 0 & \best{0.3} & TO  &  TO & 0.5 & 112 & 0.6 \\
    Solitrary   & S & \Int & 3 & 2 & 0 & \sys & 0 & \best{0.2} & 0.2 &  IR & \best{0.2} & 0.9 & 0.3 \\
    Square-5x5  & S & \Int & 4 & 2 & 4 & \sys & 0 & 0.2 & 56  &  TO & \best{0.1} & 6.8 & 0.5 \\
\hline
    Watertank Double \cite{MaderbacherB21}
                       & S & \Real& 4 & 2  & 0 & \sys & 0 & \best{0.2} & 11  & NI & NI &  TO & 0.3 \\
    Watertank Single   & B & \Real& 5 & 1  & 0 & \sys & 1 & \best{0.6} & 25  & NI &  - & 1.7 & 1.8 \\
    Elevator Simple 3  & B & \Int & 4 & 4  & 0 & \sys & 0 & \best{1.2} & 1.5 & TO &  - &  19 & 2.2 \\
    Elevator Simple 4  & B & \Int & 4 & 5  & 0 & \sys & 0 & \best{1.3} & 2.4 & MO &  - &  TO & 2.7 \\
    Elevator Simple 5  & B & \Int & 4 & 6  & 0 & \sys & 0 & \best{1.6} & 6.3 & MO &  - &  TO & 3.8 \\
    Elevator Simple 8  & B & \Int & 4 & 9  & 0 & \sys & 0 & \best{3.3} & 23  & TO &  - &  MO & 8.6 \\
    Elevator Simple 10 & B & \Int & 4 & 11 & 0 & \sys & 0 & \best{6.3} & 93  & TO &  - &  MO &  16 \\
    Elevator Signal 3  & B & \Int & 3 & 2  & 1 & \sys & 0 & \best{1.4} & 17  & MO &  - &  11 & 2.2 \\
    Elevator Signal 4  & B & \Int & 3 & 2  & 1 & \sys & 0 & \best{1.7} & 109 & MO &  - &  12 & 2.6 \\
    Elevator Signal 5  & B & \Int & 3 & 2  & 1 & \sys & 0 & \best{1.9} & TO  & MO &  - &  11 & 2.9 \\
\hline
    Example-Figure~\ref{fig:ex-motivating}
                             & R & \Int  & 2 & 1 & 1 & \sys & 1 & \best{2.1} & (IR) & (IR) & (NI) &  73 &  32 \\
    Example-Figure~\ref{fig:ex-combined}     
                             & R & \Int  & 4 & 2 & 0 & \sys & 1 & \vbest{0.7} & (TO) &   IR & (NI) &  TO & 1.2 \\
    Rob-Grid-1d-Reach        & R & \Int  & 2 & 1 & 0 & \sys & 1 & \best{0.2} & (TO) &   IR & (NI) & 2.8 & 0.7 \\
    Rob-Grid-2d-Reach        & R & \Int  & 2 & 2 & 0 & \sys & 1 & \vbest{0.3} & (MO) &   IR & (NI) &  TO & 495 \\
    Rob-Cont-1d-Reach-Real   & R & \Real & 2 & 1 & 1 & \sys & 1 & \vbest{0.3} & (TO) & (NI) & (NI) &  TO &  TO \\
    Rob-Cont-2d-Reach-Real   & R & \Real & 2 & 2 & 2 & \sys & 1 & \vbest{0.4} & (TO) & (NI) & (NI) &  TO &  TO \\
    Rob-Cont-1d-Reach-Unreal & R & \Real & 2 & 1 & 1 & \env & 0 & \best{0.1} & (TO) &   NI & (NI) & 7.5 &   - \\
    Rob-Cont-2d-Reach-Unreal & R & \Real & 2 & 2 & 2 & \env & 2 & \vbest{ 30} & (TO) &   NI & (NI) &  TO &   - \\
    Rob-Cat-1d-Real          & R & \Int  & 3 & 2 & 2 & \sys & 1 &  30 & (TO) & (MO) & (NI) & \best{9.8} &  TO \\
    Rob-Cat-2d-Real          & R & \Int  & 5 & 4 & 3 & \sys & 0 &  TO & (MO) & (MO) & (NI) &  TO &  TO \\
    Rob-Cat-1d-Unreal        & R & \Int  & 3 & 2 & 2 & \env & 1 &  31 & (TO) & (MO) & (NI) & \best{1.5} &   - \\
    Rob-Cat-2d-Unreal        & R & \Int  & 5 & 4 & 3 & \env & 0 &  TO & (TO) & (MO) & (NI) & \vbest{3.7} &   - \\
\hline
    Example-Figure~\ref{fig:ex-buchi}     
                       & B & \Int  & 5 & 2 & 1 & \sys & 4 & \best{1.9} & (MO) & (IR) &  -   & 5.3 &  TO \\
    Rob-Grid-1d-Commute& B & \Int  & 4 & 2 & 1 & \sys & 3 & \best{0.9} & (TO) &   MO &  -   &  44 &  TO \\
    Rob-Grid-2d-Commute& B & \Int  & 4 & 4 & 2 & \sys & 2 & \vbest{4.3} & (TO) &   TO &  -   &  TO &  MO \\
    Rob-Cont-1d-Commute& B & \Real & 4 & 2 & 2 & \sys & 3 & \vbest{1.4} & (TO) & (NI) &  -   &  TO &  TO \\
    Rob-Cont-2d-Commute& B & \Real & 4 & 4 & 4 & \sys & 2 & \vbest{8.4} & (MO) & (NI) &  -   &  TO &  TO \\
    Rob-Resource-1d    & B & \Int  & 4 & 2 & 1 & \env & 1 & \best{1.4} & (TO) &  IR  &  -   & 4.0 &   - \\
    Rob-Resource-2d    & B & \Int  & 4 & 3 & 2 & \env & 1 & \best{3.1} & (TO) &  MO  &  -   & 9.0 &   - \\
    Warehouse-Empty    & B & \Real, \Int 
                                  &  9 & 2 & 1 & \sys & 2 & \vbest{2.3} & (TO) & (NI) &  -   &  TO & 7.9 \\
    Warehouse-Stock    & B & \Real, \Int 
                                  & 10 & 3 & 2 & \sys & 4 & \vbest{6.1} & (MO) & (NI) &  -   &  TO &  TO \\
    Warehouse-Clean    & B & \Real, \Int 
                                  & 13 & 6 & 5 & \sys &10 & \vbest{21}  & (TO) & (NI) &  -   &  TO &  TO 
\end{tabular}}
\vspace{1mm}
\caption{%
Evaluation Results. 
$\Omega$ is the objective (\textbf{S}afety, \textbf{R}eachability, or \textbf{B}üchi). ST is the variable domain type (additional to $\Bool$).
$|L|$, $|\cells|$, $|\inputs|$ are the number of respective game elements and Win the player expected to win.
For our tool we give the number of applied accelerations \textbf{A}.
We show the wall-clock running time in seconds for our prototype \textbf{rp}gsolve, \textbf{R}aboniel, \textbf{T}eMoS, \textbf{G}enSys, and \textbf{M}uVal (with clause exchange).
We also provide the running times for our prototype with additional program \textbf{extr}action.
TO means timeout after 10 minutes, MO means out of memory (8GB), - means the approach is conceptually not applicable or extraction is not available as Player \env\ wins, NI means not-implemented (but conceptually applicable), and IR mean invalid result (by errors or incomplete refinement loops).
Parentheses $(\cdot)$ mean we expect divergence, e.g.\ due to unbounded strategy loops.
Two results are missing due to inconsitent benchmarks for reasonable comparison.
We highlight in bold the fastest solving runtime result und underline it if it is the only sucessful one.
The evaluation was performed on a single core of a Intel i7 (11thGen) processor.
}\label{tab:all}
\vspace{-8mm}
\end{table}

We implemented our game-solving method in a prototype~\footnote{Available at: \url{https://zenodo.org/doi/10.5281/zenodo.8409938}} using Z3~\cite{MouraB08} as the SMT solver.
The implementation realizes the heuristics from Section~\ref{sec:acceleration} based on the current number of symbolic state changes $k_l$ per location $l \in L$, i.e.\  how many times $a^{i+1}(l) \not\equiv_T a^{i}(l)$ so far in the computation.
Acceleration attempts incur computational costs, increasing with the templates' complexity. 
Therefore, the main goal of the heuristics is to keep the number of these attempts low and keep the used templates simple.
In Algorithm~\ref{algo:attractor-acc}, the frequency with which \textsc{Accelerate?} attempts acceleration for location $l$ grows linearly in $k_l$, as sometimes many exact steps are needed before we can successfully accelerate.
The lemma template is as described in Section~\ref{sec:resolving} and uses an invariant that has a number of conjuncts $\inv_T$ and $\inv_2$ linear in $k_l$. 
Hence, we start with simple templates, and if those are not sufficient, we use more complicated ones over time.
For $\itera$, the heuristics impose a bound on the depth of nested acceleration that is linear in $k_l$ and ensure two updates of the symbolic state per location, first an application of the enforceable predecessor and then a potential nested acceleration.
Again, the depth bound ensures that the acceleration attempts are initially simple. 
As, over time, we increase the complexity of our templates and acceleration attempts, in \resolve, we query the SMT solver with a timeout, which is quadratic in $k_l$.

We have not implemented the procedure in Figure~\ref{fig:buchi-acc} for B\"uchi acceleration in our prototype yet.
Instead, we have implemented Algorithm~\ref{algo:buchi}, in which we use $\textsc{AttractorAcc}$ instead of $\textsc{Attractor}$. 

We compare our prototype to the fixpoint engine GenSys~\cite{SamuelDK21},
the $\mu\text{CLP}$ solver MuVal~\cite{UnnoSTK20}, and
the TSL modulo theories\cite{Finkbeiner0PS19,FinkbeinerHP22} solvers Raboniel~\cite{MaderbacherB21} and TeMoS~\cite{ChoiFPS22} (run with Strix~\cite{strix} and cvc5~\cite{cvc5}).
Note that the last two perform synthesis from temporal logic specifications, and our  tool solves directly specified games.
For benchmarks where a TSL encoding was available, we use the existing benchmark, otherwise we encoded the game into TSL with an automatic translation, where the game locations are encoded with an additional data cell in the TSL formula.
For MuVal, we encoded the games into $\mu\text{CLP}$ as described in~\cite{UnnoSTK20},  where the set of  winning states is described by (nested) fixpoint equations.
We did not compare to~\cite{NeiderT16, FarzanK18} as those tools are outperformed by GenSys~\cite{SamuelDK21}, and did not compare to~\cite{MarkgrafHLNN20} as they use a fairly different model and perform similarly to GenSys on the shared benchmarks.
The implementation from~\cite{BeyeneCPR14}  is not available.
For~\cite{FaellaP23} no tool is available (instead,  a small set of SMTLib benchmark files that also contain the respective CHC encoding are available).
However, while~\cite{FaellaP23} did not compare to GenSys,  their performance seems similar.

\paragraph{Benchmarks.}
\cite{NeiderT16} introduced  a set of benchmarks that are safety games modeling one or two robots moving on an infinite two-dimensional grid while influenced by the system and the environment.
The reasoning that these games require can be localized, and no unbounded strategy loops occur.
The Cinderella game \cite{cinderella,BodlaenderHKSWZ12} is a standard safety-game benchmark for infinite-state synthesis that is parametrized in bucket size.
The number of iterations needed in a strategy is bounded.
Bloem and Maderbacher introduced a set of (TSL) benchmarks in~\cite{MaderbacherB21}, which model simple elevator and water tank controllers, and whose variables have infinite domain but take values in a bounded set.
We introduce a new set of benchmarks, described in Appendix~\ref{app:benchmarks}, where a robot moves on a grid, in a continuous space, or through a finite set of abstract locations. 
It has to perform different tasks like reaching a location, commuting between locations, executing tasks at specific locations, while handling environment disturbances, or decreasing energy levels, or avoiding an environment-controlled cat.
Our benchmarks have unbounded variable ranges and many contain unbounded strategy loops.
We did not use the TSL benchmarks from~\cite{ChoiFPS22} as most are almost deterministic (and require a manual translation from a formula into a program-like game).

\subsection{Analysis}
 Table~\ref{tab:all}~shows the results of our evaluation.
The benchmark sets from the literature ~\cite{cinderella,BodlaenderHKSWZ12,NeiderT16,MaderbacherB21} do not require acceleration, that is,  the symbolic solving procedure would terminate without acceleration.  We use them to compare our method to those from the literature.  We ran our prototype without disabling acceleration,  that is,  on all benchmarks our method attempts acceleration in accordance with the implemented heuristics.
The results in the table (column "A") show that in all cases where acceleration is not necessary, except "Watertank Single", the number of acceleration attempts that succeeded is 0.
For game solving, the evaluation demonstrates that on  standard benchmarks from the literature our method performs better than or equally well as other tools.  
More significantly,  on most of the other benchmarks,  which feature unbounded strategy loops and thus need acceleration, our prototype tool outperforms the state of the art.
Furthermore,  it also performs well when combining acceleration with explicit steps in the fixpoint computation is needed.
Except for MuVal,  none of the other tools is able to handle unbounded strategy loops.
MuVal performs well on benchmarks where the set of wining states and the required ranking arguments have concise representations.
However, it runs into scalability issues in cases where  many steps in the fixpoint computations are needed,  or when the ranking arguments become more complex.
Our prototype tool, on the other hand,  demonstrates a more consistent performance in both cases,  due to the ability of our method to combine acceleration with iterative game-solving.

For strategy extraction,  on benchmarks from the literature our prototype performs comparably to the other tools from the evaluation that support strategy extraction, which are 
Raboniel, TeMoS and GenSys. 
On our new benchmarks where acceleration is required to solve the game,  our tool is able to extract a strategy within the timeout in 5 out of 16 benchmarks (note that a strategy is only extracted for realizable problems).  
Note that, to our knowledge, no other tool is capable of extracting programs for these benchmarks, and thus our tool improves on the state of the art.
It can be seen from the results in Table~\ref{tab:all}, that when acceleration is applied,  the time our tool takes for strategy extraction is significantly higher than that for solving the game. 
The underlying reason is that the generation of strategies for acceleration requires solving relatively complex functional synthesis problems in order to synthesize Skolem functions for the acceleration lemmas.  
However, when acceleration is not needed program extraction is relatively easy, as it only amounts to keeping track of the choices enabled by the enforceable predecessors.
Our current implementation first performs Skolemization on the respective formulas and then invokes Z3 to search for a model for the Skolem function symbols.
For selected queries we experimented with syntax-guided synthesis (cvc5 with option  \texttt{--sygus-inference}) and with the AE-solver tool from~\cite{FedyukovichGG19}, but neither was able to produce a solution to any of the problem instances.
We remark that some of the $\forall\exists$ formulas contain more than 100,000 logical and arithmetic operators.
\looseness=-1

We evaluate performance of realizability checking and strategy extraction separately,  since realizability checking is of crucial importance on its own, especially in early design stages when initial specifications frequently turn out to be unrealizable.
\looseness=-1

\subsection{Discussion}
As part of future work,  we plan to further improve the scalability of our prototype tool,  especially the program extraction.
The first step would be to identify possible ways to simplify the Skolem function synthesis problem instances.  For example,  we noticed that in our problems the Skolem functions for many of the existentially quantified meta-variables are constant, and thus only for some of them  the functions involve case distinctions. If it is possible to perform some analysis at the game level to identify the different types of meta-variables,  combining different search techniques might result in simpler functional synthesis problems.

Second,  the selection of the templates and the employed generation technique for identifying acceleration lemmas are  just one way to use the general framework we propose. 
One major strength of our approach is that the needed templates capture localized arguments that are automatically combined by the acceleration procedure to reason about strategies. Thus, even if more customized user-provided templates are necessary for more complex arguments, these are still localized and do not require the user to specify details about the global strategic behavior,  which can be quite complicated.
The development of more refined lemma generation techniques, for example, inspired by Syntax Guided Synthesis~\cite{SyGuS,NiemetzPRBT21}, offers a different angle to tackle the lemma instantiation problem in future work.

A third avenue for future work is exploring the possibility to integrate in our approach solving techniques for fixpoint logics. 
Our evaluation shows that on benchmarks where the set of wining states and the required ranking arguments have concise representations,  the solver MuVal performs well. Employing similar techniques could also improve the computation of acceleration lemmas. The main challenge is designing the interface between the game-solving/acceleration procedure and the underlying reasoning method, that is, decomposing the game solving process in a way that the constraint-solving method is only required to solve simpler localized sub-problems.

At a higher level,  another direction is the investigation and integration of approximation techniques. Our current approach computes exact winning sets, which might need complex representations, while in many cases computing approximations might be sufficient. 
Further directions include exploring the relation of reactive program games to temporal logics.

\section{Conclusion}\label{sec:conclusion}
We study reactive program games, a type of infinite state games with temporal objectives.
We propose a symbolic method for solving such games that relies on a novel technique for accelerating the game-solving process in the presence of unbounded strategy loops. 
This acceleration is the key reason why our method can solve games on which state-of-the-art techniques diverge, as we demonstrate in the evaluation of our prototype implementation. 
In this paper we focus on infinite-state games with safety, reachabiliy, B\"uchi and co-Büchi objectives.  We believe that by integrating the accelerated attractor computation in symbolic fixpoint-based algorithms for parity games, we can obtain an acceleration-enhanced method for parity games.  The evaluation  of acceleration for solving parity games and the development of  dedicated acceleration methods is a topic of future work.
Since our acceleration method's core idea is based on generic inductive statements, we believe our work expands the scope of infinite-state synthesis and opens up a range of interesting directions for further development.

\bibliographystyle{ACM-Reference-Format}
\bibliography{main.bib} 

\appendix

\section{Proofs}\label{app:proofs}

 Theorem~\ref{thm:attractor-acc}
\emph{
Let $\mathcal{G}$ be a reactive program game structure,   $p \in \{\sys,\env\}$,  and $d \in \symstates$.\\
If $\textsc{AttractorAcc}(\mathcal G, p,d)$ terminates returning $a$, then it holds that $\sema{a} = \mathit{Attr}_{\sema{\mathcal G},p}(\sema{d})$. 
}
\begin{proof}
The overall correctness of Algorithm~\ref*{algo:attractor-acc} follows from the soundness of the fixed-point computation, as already stated in Proposition~\ref{prop:correctness-attractor}.
This follows from the correctness of the classical attractor algorithm.
The only remaining part to show is that for every $n \geq 0$, 
\[\resolve(\accreach(\mathcal{G}, p, l, a^n)) \subseteq \mathit{Attr}_{\sema{\mathcal G},p}(\sema{d}).\]
Assume that $\accreach$ returns $(\constr, \varphi) := \accreach(\mathcal{G}, p, l, a^n)$. 

If function $\resolve{}$ finds some map $\lambda$ such that the formula $\constr[\lambda]$ is valid,
it returns the corresponding formula $\varphi[\lambda]$.
By Lemma~\ref{lem:correcness-accA},  we have that
\[\sema{\{ l \mapsto \varphi[\lambda]\}} \subseteq \mathit{Attr}_{\sema{\mathcal G},p}(\sema{a^n}).\]
As $\sema{a^n} \subseteq \mathit{Attr}_{\sema{\mathcal G},p}(\sema{d})$,  it follows that the property holds.

If, on the other hand, $\resolve{}$ does not find a map $\lambda$ it returns $\bot$, and hence,  the inclusion  in $\mathit{Attr}_{\sema{\mathcal G},p}(\sema{d})$ trivially holds.
\end{proof}

\bigskip 

 Lemma~\ref{lem:correcness-accA}
\emph{
Let $\mathcal{G}$ be a reactive program game structure,  $p \in \{\sys,\env\}$,  $l \in L$ and $d \in \symstates$. 
Then, if $\accreach(\mathcal G, p,l,d)$ returns the pair $(\constr,\varphi)$, then for every mapping $\lambda: \usedlemmas(\{\constr, \varphi\}) \to \lemmas(\cells)$ such that the formula $\constr[\lambda]$ is valid it holds that $\sema{\{ l \mapsto \varphi[\lambda]\}} \subseteq \mathit{Attr}_{\sema{\mathcal G},p}(\sema{d})$.}
\begin{proof}
We prove this statement via structural induction over the recursion tree of $\accreach$.
Note that this recursion tree is well-founded as we assume $\accreach$ to terminate.
Since both the base case and induction step are fairly similar, we will make the case distinction later on in the proof.
We denote with $\base$, $\step$, $\conc$ the instantiations of $b$, $s$, $c$ by $\lambda$, respectively.
Note that $\Inv$ and $d$ do not contain any of the uninterpreted symbols introduced by the symbolic acceleration.
If the formula $\constr[\lambda]$ is valid modulo the theory $T$,  so are the following sub-statements .
\begin{itemize}
    \item[(1)] $\forall \cells.~\Inv(l) \land \base \to d(l)$
    \item[(2)] $\forall \cells.~\Inv(l) \land \lnot d(l) \land \conc \to a[\lambda](l)$
    \item[(3)] $\forall \cells.~\Inv(l) \land \lnot d(l) \land \conc \to \constr_\mathit{Rec}[\lambda]$
\end{itemize}

Depending on the state of the induction, we conclude the following:
\begin{itemize}
    \item[{\bf Base Case:}]
        Since we are at a leaf of the recursion tree, $\itera$ does not call $\accreach$ again.
        Hence, $\itera$ is by construction an under-approximation of the attractor computation.
        This means for some assignment $\assmt{x}\in \assignments{\cells}$ substituted for the respective $E$ it holds that
        \[ \sema{a[\lambda]} \subseteq \attr{\sema{\mathcal{G}_\mathit{loop}}, p}{\sema{d} \cup \{\locend\} \times \{\assmt{x}' \in \assignments{\cells}  \mid \combine{\assmt{x}}{\assmt{x}'} \FOLentailsT{T} \step \})}, \]
        where the combination function $\combine{\cdot}{\cdot}$ is taken from Definition~\ref{def:lemma}.
    \item[{\bf  Induction Step:}]
        By induction hypothesis we know that for every assignment $\assmt{x} \in \assignments{\cells}$ substituted for the respective $E$, if $\constr_\mathit{Rec}[\lambda]$ is valid, then all cursive calls of $\accreach$ also compute a subset of the attractor (as all their constraints have to be valid) of $d_\mathit{loop}$ and hence $\itera$ computes a subset of the attractor.
        Similar to above, in this case we have that
        \[ \sema{a[\lambda]} \subseteq \attr{\sema{\mathcal{G}_\mathit{loop}}, p}{\sema{d} \cup \{\locend\} \times \{\assmt{x}'\in  \assignments{\cells} \mid \combine{\assmt{x}}{\assmt{x}'} \FOLentailsT{T} \step \})} .\]
        Note that by the validity of (3),   $\constr_\mathit{Rec}$ is valid, if $\assmt{x} \FOLentailsT{T} \Inv(l)$, $\assmt{x} \FOLentailsT{T} \lnot d(l)$, $\assmt{x} \FOLentailsT{T} \conc$.
\end{itemize}

Hence, there exists a strategy $\sigma_S$ for player $p$ in $\mathcal{G}_\mathit{loop}$ with the following property:
For all $\assmt{y} \in \assignments{\cells}$ with $\assmt{y} \FOLentailsT{T} \Inv(l), \lnot d(l), \conc$, it holds that 
$\assmt{y} \FOLentailsT{T} a[\lambda](l)$ (by condition (2)). 
Hence, for all $\pi \in \plays((l, \assmt{y}) , \sigma_I)$, $\pi$ reaches $\sema{d}$ or 
$\{\locend\} \times \{\assmt{y}'  \in \assignments{\cells} \mid \combine{\assmt{y}}{\assmt{y}'} \FOLentailsT{T} \step \}$.

Note that the uninterpreted constants in $E$ have been mapped to $\assmt{y}$,  since for $a[\lambda](l)$ the constants $E$ have been mapped to $\cells$ which are set by the assignment $\assmt{y}$.
This also shows why we have to be careful with the quantifiers as we map the uninterpreted  constants in $E$ to the variables $\cells$ syntactically, and now, the variables $\cells$ are mapped semantically to the respective values in the assignment $\assmt{y}$.
For this to be correct, it is necessary that the variables in $\cells$ to which the constants $E$ are mapped are free variables (i.e. ,  are not bound by some nested quantifier).

We now explain how we can construct a strategy $\sigma_I$ for player $p$ in $\mathcal{G}$ that iterates the strategy $\sigma_S$ (in the loop game $\mathcal{G}_\mathit{loop}$) in order to enforce reaching $\sema{d}$ in $\mathcal{G}$ starting from an assignment $\assmt{x}$.
Initially, the strategy $\sigma_I$ simulates the moves of $\sigma_S$.  
If $\sema{d}$ is reached,  $\sigma_I$ can behave arbitrarily from that point on, as it has reached the goal.
Otherwise, we can reach $\locend$ by $\sigma_I$, i.e. , reach the set of states
$\{l\} \times \{\assmt{x}' \in \assignments{\cells} \mid \combine{\assmt{x}}{\assmt{x}'} \FOLentailsT{T} \step \}$.
From there,  $\sigma_I$ restarts the strategy $\sigma_S$ from location $l$, unless the current state is in $\sema{d}$.
Note that this is possible, since the current state belongs to $\Inv(l) \land \lnot d(l) \land \conc$.
This is the case because the strategies stay in the game (i.e.\ in $\Inv(l)$) and by Definition~\ref{def:lemma} 
$\{\assmt{x}' \in \assignments{\cells} \mid \combine{\assmt{x}}{\assmt{x}'} \FOLentailsT{T} \step \} \subseteq  \{ \assmt{z} \in \assignments{\cells} \mid \assmt{z} \FOLentailsT{T} \conc \}$. 

We now show that $\sigma_I$ guarantees that $\sema{d}$ is reached for all possible behaviors of the opponent player. 
Assume towards a contradiction that this is not the case.
This means that by iterating the above property of $\sigma_I$ we can constrict an infinite sequence $\alpha \in \assignments{\cells}^\omega$ of program variable assignments that we visit while looping though $l$.
This sequence starts in $\assmt{x}$,  and for all $i \in \Nat$,  it holds that $\combine{\alpha[i]} {\alpha[i+1]} \FOLentailsT{T} \step$.
However, Definition~\ref{def:lemma} implies that there exists $k \in \Nat$, such that $\alpha[k] \FOLentailsT{T} \base$.
Since the strategy cannot leave the game,  this means that $\alpha[k] \FOLentailsT{T} Inv(l)$ and hence,  by condition (1), $\alpha[k] \FOLentailsT{T} d(l)$ which yields a contradiction.

Thus,   we showed that starting from any assignment in $\{ \assmt{x} \in \assignments{\cells} \mid \assmt{x} \FOLentailsT{T} \Inv(l) \land \conc \land \lnot d(l)\}$  the strategy $\sigma_I$ will enforce that  $\sema{d}$ is reached. Hence,  we can conclude that in location $l$,  the set of assignments $\{ \assmt{x} \in \assignments{\cells} \mid \assmt{x} \FOLentailsT{T} \Inv(l) \land \conc \}$ is a subset of $\mathit{Attr}_{\sema{\mathcal G},p}(\sema{d})$. 
This completes the proof of our statement.
\end{proof}

\section{Benchmarks}\label{app:benchmarks}
All our \textbf{Rob}-* benchmarks describe a moving robot that can move one step in one direction (positive or negative, x-axis or y-axis direction).
It cannot move in the diagonal direction.
We refer by the origin the location $x = 0$ or $x = 0 \land y = 0$, depending on the dimension.
\textbf{Rob-Grid-1d-Reach} is a reachability game where the robot has to reach the origin on a one-dimensional integer-grid from some arbitrary initial location.
\textbf{Rob-Grid-2d-Reach} is the same as the previous benchmark, but in two dimensions.
\textbf{Rob-Grid-1d-Commute} is a Büchi game where the robot has to commute (i.e.\ move back and forth) forever between the origin and some target location (picked after every commute by the environment) on a one-dimensional integer grid.
\textbf{Rob-Grid-2d-Commute} is the same as the previous benchmark, but in two dimensions.
\textbf{Rob-Cont-1d-Reach-Real} is a reachability game where the robot has to reach the origin in a one-dimensional real space from some arbitrary initial location. 
In addition, the environment inserts some noise into each movement of the robot.
The noise's absolute value is smaller than $0.3$.
\textbf{Rob-Cont-2d-Reach-Real} is the same as the previous benchmark, but in two dimensions.
\textbf{Rob-Cont-1d-Reach-Unreal} is the same as \textbf{Rob-Cont-1d-Reach-Real} but with a noise of $1.3$ which renders the problem winning for the environment.
\textbf{Rob-Cont-2d-Reach-Unreal} is the same as the previous benchmark, but in two dimensions.
\textbf{Rob-Cont-1d-Commute} is the same as \textbf{Rob-Grid-1d-Commute} but in the continuous space with environment noise instead of the grid.
\textbf{Rob-Cont-2d-Commute} is the same as the previous benchmark, but in two dimensions.
In \textbf{Rob-Cat-1d-Real},
the robot moves on the one-dimensional integer grid and has to reach the origin.
However, it has to avoid getting caught by the cat, which moves like the robot.
To make the game winning for the system, the cats start farther away from the origin than the robot.
\textbf{Rob-Cat-2d-Real} is the same as the previous benchmark, but in two dimensions.
\textbf{Rob-Cat-1d-Unreal} is the same as \textbf{Rob-Cat-1d-Real}, but the cat can start wherever it wants, making it winning for the environment.
\textbf{Rob-Cat-2d-Unreal} is the same as the previous benchmark, but in two dimensions.
In \textbf{Rob-Resource-1d}, the robot moves on the one-dimensional integer grid and has to visit a target location that is provided by the environment.
Once it reaches it, it looses one resource and gets an new target.
It has to do this over and over again and looses if it the resources fall below zero. 
However, it starts with only four resources.
\textbf{Rob-Resource-2d} is the same as the previous benchmark, but in two dimensions.

The warehouse benchmark models a warehouse with four floors on which the robot can be. 
Each floor might have attached special locations. 
The robot can stay or move to the next and previous floor or to the currents floor special location. 
The robot has a real energy level which should not fall below zero, otherwise the system looses.
Each move the robot looses an environment defined amount of energy between zero and one units.
On floor zero, there is the charging special location where the robot can recharge its energy by 1.5 per move and leave if it wants.
Also on floor zero, there is the base station where the robot awaits further tasks.
However, while doing nothing it will also still loose energy.
This location has to be visited infinitely often.
On floor two, there is a trap the robot cannot leave anymore.
Note that the choices of the robot are encoded using integers.
In \textbf{Warehouse-Empty}, the robot has to do nothing additional than not loosing to much energy.
In \textbf{Warehouse-Clean}, while idling the robot might get environment issued cleaning orders for the different floors. 
It then has to go cleaning before it goes back to idling again.
In \textbf{Warehouse-Stock}, the robot is not allowed to idle but has to restock up to four items infinitely often.

\end{document}